\documentclass[%
 reprint,
 amsmath,amssymb,
 aps,
prx
]{revtex4-2}

\usepackage{graphicx}
\usepackage{dcolumn}
\usepackage{bm}

\usepackage{xcolor}
\begin{document}

\title{Interplay between dressed and strong-axial-field states in Nitrogen-Vacancy centers for quantum sensing and computation 
}

\author{G. Zanelli$^{1,2}$}
\author{E. Moreva$^1$}
\author{E. Bernardi$^1$}
\email{e.bernardi@inrim.it}
\author{E. Losero$^1$}
\author{S. Ditalia Tchernij$^{2,3}$}
\author{J. Forneris$^{2,3}$}
\author{\v{Z}. Pastuovi\'{c}$^4$}
\author{P. Traina$^1$}
\author{I. P. Degiovanni$^1$}
\author{M. Genovese$^{1,3}$}

\affiliation{$^1$Istituto Nazionale di Ricerca Metrologica (INRiM), Strada delle cacce 91, Torino, Italy}
\affiliation{$^2$Physics Department and NIS Centre of Excellence - University of Torino, Torino, Italy}
\affiliation{$^3$Istituto Nazionale di Fisica Nucleare (INFN) Sez. Torino, Torino, Italy}
\affiliation{$^4$Centre for Accelerator Science, Australian Nuclear Science and Technology Organisation, New Illawarra rd., Lucas Heights, NSW 2234, Australia}

\begin{abstract}

 The Nitrogen-Vacancy (NV) center in diamond is an intriguing electronic spin system with applications in quantum radiometry, sensing and computation. In those experiments, a bias magnetic field is commonly applied along the NV symmetry axis to eliminate the triplet ground state manifold's degeneracy (S=1). In this configuration, the eigenvectors of the NV spin's projection along its axis are called strong-axial field states.
 Conversely, in some experiments, a weak magnetic field is applied orthogonally to the NV symmetry axis, leading to eigenstates that are balanced linear superpositions of strong-axial field states, referred to as dressed states. The latter are sensitive to environmental magnetic noise at the second order, allowing to perform magnetic field protected measurements while providing increased coherence times. However, if a small axial magnetic field is added in this regime, the linear superposition of strong-axial field states becomes unbalanced. 
This paper presents a comprehensive study of Free Induction Decay (FID) measurements performed on an NV center ensemble in the presence of strain and weak orthogonal magnetic field, as a function of a small magnetic field applied along the NV symmetry axis. The simultaneous detection of dressed states and unbalanced superpositions of strong-axial field states in a single FID measurement is shown, gaining insight into coherence time, nuclear spin, and the interplay between temperature and magnetic field sensitivity.
The discussion concludes by describing how the simultaneous presence of magnetically-sensitive and -insensitive states opens up appealing possibilities for both sensing and quantum computation applications.

\end{abstract}


\maketitle

\newcommand{\ket}[1]{\mbox{\ensuremath{|#1\rangle}}}
\newcommand{\bra}[1]{\mbox{\ensuremath{\langle#1|}}}


\section{Introduction} 
The use of quantum states to sense physical observables, known as quantum sensing \cite{cap2017rev,Petrini2020biosens}, has demonstrated to be of factual advancement in a plethora of diverse applications (spanning from magnetometry, gravitational wave detection and quantum thermodynamics)\cite{oh2024qsens,Genovese2021interf,Karsa2024radar,GALL2024radar,Brida2011twbeams,YON2014NOON,Aas2013LIGO, str2022grav,Mag2022RF,Cappellaro2024Thermo}. However, isolating a sensor from different sources of noise represents a major challenge in high-precision quantum sensing. Among several classes of quantum sensors, the Nitrogen-Vacancy (NV) center in diamond has been extensively studied as a high-sensitivity nanoscale sensor.  More precisely, the NV is an optically active point defect composed of a single substitutional Nitrogen and a lattice vacancy in nearest-neighbour configuration; this structure leads to an electronic spin triplet configuration S=1, which can be exploited via spin resonance techniques, allowing for the so-called Optically Detected Magnetic Resonance (ODMR)\cite{doherty2012theory} (for an introductory description of NV levels structure and ODMR principle see \cite{doherty2013nitrogen}). NV-based sensing is commonly carried out by monitoring the effect of environmental variables on the dynamics of the NV center, this scheme is enabled by three main features \cite{rondin2014magnetometry}: (i) the spin state can be controlled using microwave radiation; (ii) the initialization and readout can be performed optically; (iii) the system presents coherence times $T_{2}^{*}$ in the order of tens of microseconds at room temperature\cite{doherty2013nitrogen, ofori2012spin}.\\Various techniques have been introduced to make measurements selective on single physical observables, i.e., temperature or magnetic field. For example, thermal echo \cite{toyli2013fluorescence}, D-Ramsey pulse sequences \cite{neumann2013high}, quantum beats magnetometry \cite{fang2013high}, multipulse dynamical decoupling \cite{Pham2012DDec} and spectral hole burning \cite{kehayias2014microwave} have been explored.
Nonetheless, as it will be detailed, each environmental parameter may act differently depending on the quantum state of the NV center, thus a thorough analysis of the competition of these parameters in the involved physical process is required. This study will focus on Continuous-wave (CW)ODMR and Free Induction Decay (FID) sensing protocols; CW-sensing protocols aim to detect the variation induced by external fields on the resonance frequency of the NV center. Free Induction Decay sensing protocols measure the phase difference acquired as a consequence of variation in resonance frequency between two components of a quantum superposition. Sensors based on NV centers present a very good sensitivity regarding temperature and magnetic field \cite{barry2020sensitivity}, leading to the possibility of measuring the temperature inside a cell \cite{kucsko2013nanometre,petrini2022nanodiamond,fujiwara2020real}, magnetic NMR signal coming from a single molecule \cite{lovchinsky2016nuclear, glenn2018high} as well as nanoscale imaging of superconducting vortexes \cite{thiel2016quantitative}. Additionally, the NV is widely used in computation applications where paramagnetic $^{13}$C nuclei are employed as q-bits, the NV center as a mediator q-bit, and the neighboring N nucleus as auxiliary q-bit \cite{abobeih2022fault, bradley2019ten}. Moreover, In recent applications, the possibility of using the network of $^{13}$C nuclei for quantum simulation has been investigated \cite{van2024mapping}.\\In most of these applications, a bias magnetic field $B_{\parallel}$ is applied along the NV axis to remove the degeneration of the $S_z=\pm1$ manifold. In this case, the eigenstates of the system are NV's axial spin eigenstates, namely $\ket{S_z= +1}$ and $\ket{S_z=-1}$, generally indicated as strong-axial field states. Conversely, 
in the presence of a weak magnetic field, electric field or strain (or a combination) orthogonal to the NV axis 
(with $B_{\parallel}=0$), the eigenstates are balanced linear superposition of the strong-axial field states: $\ket{+} = \frac{1}{\sqrt{2}}\left(\ket{S_z=+1} + \ket{S_z=-1}\right)$ and $\ket{-} = \frac{1}{\sqrt{2}}\left(\ket{S_z=+1} - \ket{S_z=-1}\right)$, typically called dressed states \cite{doherty2012theory}.
\\Dressed states have proven to be useful for temperature \cite{moreva2020practical} and electric field measurements \cite{chen2017high} since they show a second-order dependency to the magnetic field due to the null spin expectation value for each direction: $\bra{+}S_i\ket{+}=\bra{-}S_i\ket{-}=0$ for $i=x,y,z$, leading to longer coherence times \cite{shin2013suppression} when the dominant source of decoherence is magnetic noise \cite{jamonneau2016competition}.
\\Dressed states have been studied both for single NV center\cite{dolde2011electric, jamonneau2016competition, rao2020level} and NV ensemble \cite{clevenson2016diamond, shin2013suppression, lamba2024vector, chen2017high}. In the last few years, research has been focusing on the effect of $^{14}$N \cite{clevenson2016diamond} and $^{13}$C \cite{rao2020level} nuclei on the NV center and their exploitation for electrical field sensing\cite{qiu2022nanoscale, chen2017high}. In particular, in the presence of a weak orthogonal field ($B_{\bot}\neq0$, $B_{\parallel}=0$), it has been shown that two resonances are present for both $\ket{-}$ and $\ket{+}$\cite{clevenson2016diamond}. The first is related to the state $I^{^{14}N}_z=0$, and the other is related to the two degenerate states $I^{^{14}N}_z=\pm1$, where $I^{^{14}N}_z$ is the component of the $^{14}$N nuclear spin along the NV axis. This result proves that the NV center interacts with the $^{14}$N nucleus despite being almost insensitive to external magnetic fields. The reason is that, for $B_\parallel=0)$, the state with $I^{^{14}N}_z=0$ is a balanced superposition of strong-axial field states, therefore it is a completely dressed state. The states with  $I^{^{14}N}_z=+1$ and $I^{^{14}N}_z=-1$, instead, are unbalanced superpositions of the strong-axial-field states with $\langle S_z \rangle>0$ and $\langle S_z \rangle <0$ respectively, as a result they are more sensitive to the external magnetic field. In this work, these unbalanced superpositions of $\ket{S_z=+1}$ and $\ket{S_z=-1}$ are named partially-dressed states. Finally, it should be stressed that applying a weak orthogonal field to observe the interaction between dressed and partially dressed states in the FID decays is unnecessary. The total electric field alone has a similar effect on the eigenenergies of a single NV\cite{wang2022zero} and is likely to have similar effects on the FID decays. However, employing a weak orthogonal field can enhance the stability of the dressed states by increasing the energy gap, when properly aligned. Additionally, it eliminates the overlap of resonances associated with different orientations of NV's when working with NVs ensembles.\\
In this paper, via a comprehensive theoretical and experimental investigation of the interplay between dressed and strong-axial field states in the presence of both a weak orthogonal magnetic field and an electric field (or, equivalently, a strain), it is demonstrated, for the first time, the possibility of simultaneously exciting dressed and partially-dressed states. Furthermore, we present a novel approach consisting of fitting single FID measurements with multiple coherence times, that is of critical importance when working with different classes of states. This methodology will lead to the implementation of two-qubit gates with increased fidelity due to the increased coherence time $T_2^*$ of dressed and partially-dressed states. Moreover, it could enable a decoupled magnetic field and temperature sensing scheme, operating with limited microwave bandwidth and with a single microwave frequency. In the first part of the work, how strain and weak orthogonal fields compete in forming dressed states is discussed, providing, as a result, the exact eigenstates of the system. Then, FID measurements obtained for two different values of the axial magnetic field are presented, highlighting the presence of dressed and partially-dressed states. The study is focused on NV ensembles, but it can be straightforwardly generalized to the single NV center or other solid-state spin systems.

\section{Theoretical analysis}
The system considered is constituted of NV center electronic spin $\vec{S}=1$ and $^{14}$N nuclear spin $\vec{I}=1$. The complete Hamiltonian of this system is \cite{doherty2012theory}:

\begin{equation}
\label{eq:tot_ham}
\begin{split}
& \mathcal{H}= (D_{gs}+ d_{\parallel}\Pi_{\parallel})\left[S_z^2-\frac{1}{3}S(S+1)\right] \\
&- d_{\bot}\left[\Pi_x(S_x^2- S_y^2)-\Pi_y(S_xS_y + S_yS_x)\right] \\
&+ g_e\mu_BB_xS_x + g_e\mu_BB_yS_y\\ 
&+  g_e\mu_BB_{\parallel}S_z\\
&+ S_zA_{\parallel}I_z + S_xA_{\bot}I_x + S_yA_{\bot}I_y\\
&+Q \left( I_z^2-\frac{I^2}{3}\right)\\
&+ g_n\mu_nB_xI_x + g_n\mu_nB_yI_y\\
&+g_n\mu_n{B_{\parallel}}I_z,
\end{split}
\end{equation}

\begin{figure*}[ht!]
\includegraphics[width=\textwidth]{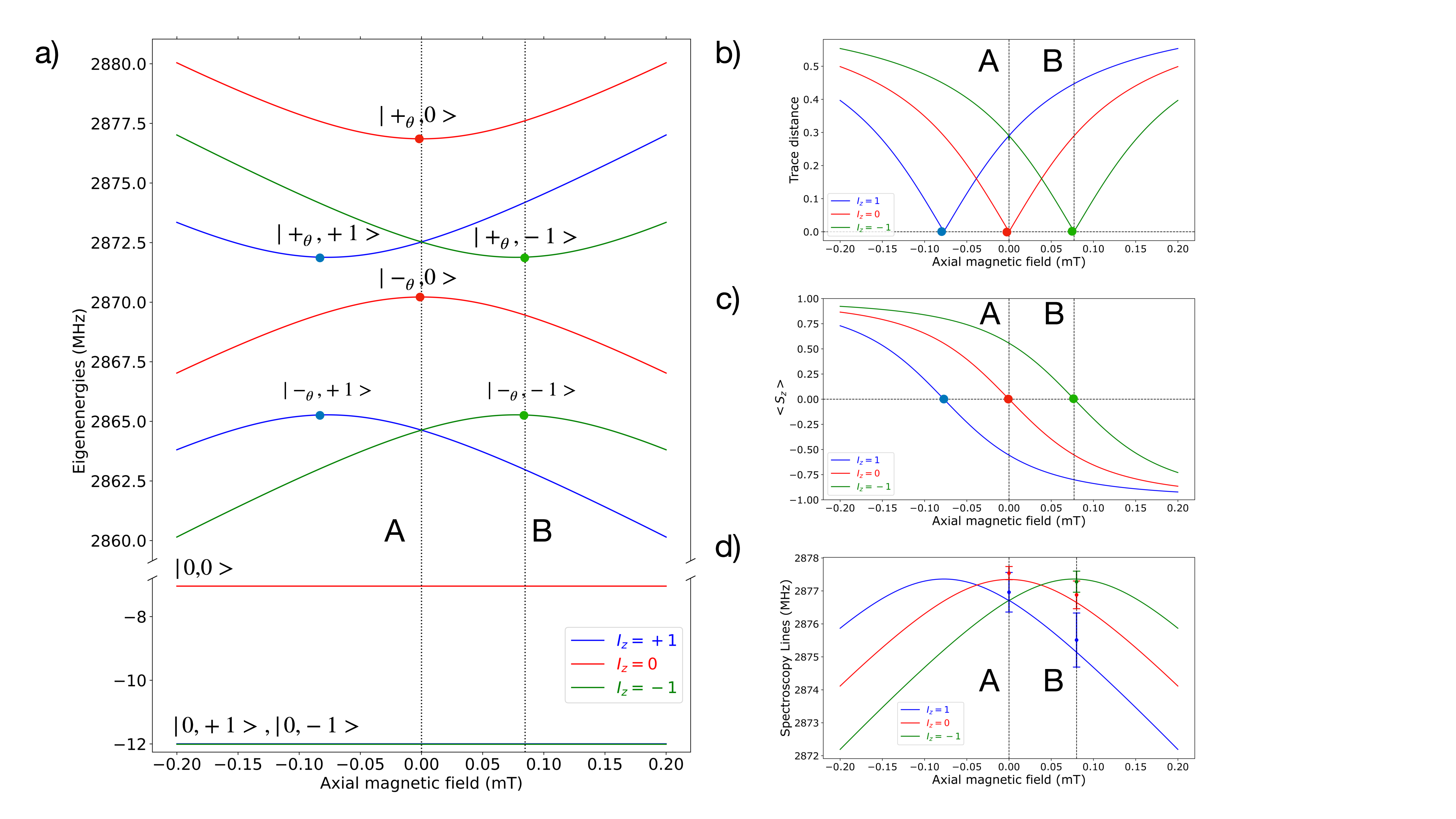}
\caption{(a) Exact eigenvalues of Hamiltonian in Eq. (\ref{eq:tot_ham}) as a function of axial magnetic field $B_{\parallel}$ ($B_{\bot}$ fixed). The curves are labeled according to the value of the spin eigenstates \ket{S_z, I_z},  where $S_z$  and $I_z$ are the axial components of electronic spin and nuclear spin, respectively. In particular, the position of the dressed states \ket{+_{\theta} ,\cdot} and \ket{-_{\theta} ,\cdot} are depicted as small dots. (b) Trace distance between the density matrix derived from electronic dressed state $\ket{-}_{\theta}$ and the density matrix derived from the eigenvectors of Eq. (\ref{eq:tot_ham}), as a function of the axial magnetic field $B_{\parallel}$. (c) Expectation values of $S_z$ on the dressed states eigenvectors \ket{-_{\theta} ,\cdot}. (d) Resonance frequencies from $\ket{0}$ to the lower branch of excited states as a function of axial magnetic field $B_{\parallel}$. The energy differences are computed taking into account selection rules (the applied microwave couples with $S_x$ and $S_y$, therefore $\Delta I_z=0$, $\Delta S_z\pm1$.) The circles represent the resonance frequencies derived from cw-ODMR spectra, see Figs.\ref{fig:expA}a), \ref{fig:expB}a). The uncertainties in experimental data are the line widths of the respective ODMR resonances.}
\label{fig:energytheo}
\end{figure*}

where $D_{gs}$ is the ground state Zero Field splitting, $d_{\bot}$ and $d_{\parallel}$ are the components of the ground state electric dipole moment, $\vec{\Pi}=\vec{E} + \vec{\sigma}$ is the total effective electric field that includes both the effect of static electric fields $\vec{E}$ and strain $\vec{\sigma}$ (strain acts as an effective electric field due to the piezoelectric effect). $g_e$ and $g_n$ are electronic and nuclear Land\'{e} $g$ factors, $\mu_B$ and $\mu_N$ are the Bohr and nuclear magneton constants, $\vec{B}$ is the applied magnetic field, $A_{\parallel}$ and $A_{\bot}$ describe the hyperfine interaction between $\vec{S}$ and $\vec{I}$, and $Q$ is the nuclear electric quadrupole term.

In Fig. \ref{fig:energytheo} (a), the eigenvalues of Hamiltonian \cite{johansson2012qutip} in Eq. (\ref{eq:tot_ham}) are numerically evaluated \cite{qutip1} in the presence of a weak fixed orthogonal field $\vec{B_{\bot}}$ and plotted functions of a varying axial magnetic field $B_{\parallel}$. This orthogonal field has the form $\vec{B_{\bot}} = B_x\hat{x}+B_y\hat{y}$, with $B_x= 3.83$ mT and$B_y=3.33$ mT. Additionally, an orthogonal total electric field $\vec{\Pi_{\bot}} = \Pi_x\hat{x}+\Pi_y\hat{y}$ is considered, with $\Pi_x=-124000$ V  cm$^{-1}$ and $\Pi_y=-94000$ V  cm$^{-1}$ (the values reported are obtained to match the experimental data, see Supplemental Material).   

In the following, the effect of each term in Eq. (\ref{eq:tot_ham}) is analyzed to clarify the nature of the eigenstates, first by describing the effects of electronic interaction terms, and successively by including the nuclear interaction terms also.

\subsubsection{Electronic Interaction terms }
We start by considering the case where the nuclear spin interaction terms are neglected. This choice allows to limit the analysis to electron-related terms of the Hamiltonian only, as in ref. \cite{doherty2012theory}. There, the unperturbed Hamiltonian depends on the ground state crystal field splitting $D_{gs}$ while the contributions from the orthogonal total electrical field $\vec{\Pi_{\bot}}$ and the orthogonal field $\vec{B_{\bot}}$ are considered as perturbations, developing the perturbative solution up to the second order in eigenenergies (the effect of parallel strain $\Pi_{\parallel}$ is negligible because $d_{\parallel}\ll d_{\bot}$). In the present analysis, instead, the exact eigenstates are computed by diagonalization of the partial Hamiltonian containing only the electronic spin terms (zero-field, magnetic field, electric field). Then, the newly obtained eigenstates and eigenenergies are expressed as a truncated power series of the parameter $\zeta=\frac{g_{e}\mu_{B}B_{\bot}}{D_{gs}}$( for detailed calculations see Appendix \ref{Calculations eigenvalues}).

The effect of $D_{gs}$ is to create a large gap between the eigenenergy $E_{\ket{S_z=0}}$, relative to the eigenstate $\ket{S_z=0}$,  and the eigenenergies $E_{\ket{-}}=E_{\ket{+}}$ relative to the two degenerate states $\ket{-}$ and $\ket{+}$. The combined contribution of the total orthogonal electric field $\vec{\Pi_{\bot}}$  and weak orthogonal field $\vec{B_{\bot}}$ can be summarized in four effects:

\begin{itemize}
	\item Rotating the original dressed states  $\ket{0},\ket{-},\ket{+}$ of an angle $\theta$ along the $z$-axis, giving the eigenstates $\ket{0}_{\theta},\ket{-}_{\theta}, \ket{+}_{\theta}$. With

\begin{equation}
\label{eq:theta}
\theta = \frac{1}{2}\arg\left(e^{2i \phi _{B_{\bot}}}-2R\cdot e^{-i \phi _{\Pi}}\right)
\end{equation}
	
	\item Creating an energy gap between $\ket{-}_{\theta}$ and $\ket{+}_{\theta}$, where:

\begin{equation}
\label{eq:eigenenergies}
E_{gap}=2\left[ \left(\frac{\mathcal{B}_{\bot}^2 }{2D_{gs}}\right)^2 + \mathcal{E}^2 -  \mathcal{E}\frac{\mathcal{B}_{\bot}^2 }{D_{gs}}\cos(2 \phi _{B_{\bot}} +\phi_{\Pi}) \right]^\frac{1}{2}
\end{equation}
	
	\item Creating a small mixing, proportional to $\frac{\mathcal{B}}{D_{gs}}$, between $\ket{+}_{\theta},\ket{-}_{\theta}$ and $\ket{0}$
	\item Decreasing the energy of the $\ket{0}$ state to $E_0= - \frac{\mathcal{B}_{\bot}^2}{D_{gs}}< 0$
\end{itemize}

Where the following quantities are introduced:

\begin{equation}
\label{eq:parameters}
\mathcal{B}_{\bot}= g\mu_BB_{\bot}  \,\,,\,\,  \mathcal{E}=d_{\bot}\Pi_{\bot} \,\,,\,\,  R=\frac{\mathcal{E}D_{gs}}{\mathcal{B}_{\bot}^2},
\end{equation}

and

\begin{equation}
\label{eq:angles}
 \phi_{\Pi}=\arctan\left(\frac{\Pi_y}{\Pi_x} \right)    \,\,,\,\, \phi_{B_{\bot}}=\arctan\left(\frac{B_y}{B_x} \right)
\end{equation}


The explicit expression of the rotated eigenstates is (for the detailed theoretical model, see Appendix \ref{Calculations eigenvalues}):

\begin{equation}
\label{eq:eigenvectors_rotated}
\begin{split} 
& \ket{0}_{\theta} = \ket{0} \\
& \ket{-}_{\theta} = \frac{1}{\sqrt{2}}\left(e^{-i\theta}\ket{S_z=+1}- e^{i\theta} \ket{S_z=-1}\right)\\
& \ket{+}_{\theta} = \frac{1}{\sqrt{2}}\left(e^{-i\theta}\ket{S_z=+1} + e^{i\theta} \ket{S_z=-1}\right)\\
\end{split}
\end{equation}

It is significant to study the dependency between $\theta$ and $R$, which are the rotation angle of dressed states and the relative magnitude of $\Pi_{\bot}$ and $B_{\bot}$ respectively, since the latter is defining how dressed states are aligned with respect to bias electric and magnetic fields. First, we consider the two limit cases for $R$. On the one hand, when the contribution of the orthogonal magnetic field $B_{\bot}$ is dominant ($R\sim 0$) the eigenstates are rotated by $\phi_{B_{\bot}}$, i.e. are aligned along the direction of  $B_{\bot}$. On the other hand, when the contribution from the total orthogonal electric field $\Pi$ is dominant ($R \gg 1$), the original eigenstates $\ket{0}, \ket{-}, \ket{+}$ are rotated by $\frac{\pi}{2} -\phi_{\Pi}/2$. For intermediate values of $R$, the competition between the orthogonal magnetic field $B_{\bot}$ and the orthogonal total electric field $\Pi_{\bot}$ results in a intermediate rotation between $\phi_{B_{\bot}}$ and $\frac{\pi}{2} -\phi_{\Pi}/2$, determined by the relative magnitude of $B_{\bot}$ and $\Pi_{\bot}$. The general expression of $\theta$ shown in Eq. (\ref{eq:theta}) represents the first novel result of this paper.\\
The eigenvectors in Eq.s (\ref{eq:eigenvectors_rotated}) are still dressed states, i.e., a balanced superposition of strong-axial field states, thus being sensitive to the field $\vec{B}$ only at second order since:

\begin{equation}
\begin{split}
& _{\theta}\langle S_z=0 | S_i | S_z=0  \rangle_{\theta} = _{\theta}\langle - | S_i | - \rangle_{\theta}= _{\theta}\langle + | S_i | + \rangle_{\theta} =0 \\
& \text{for}  \, i=x,y,z.
\end{split}
\end{equation}

By introducing a small axial magnetic field ($\mathcal{B}_{\parallel}^2 < E_{gap}/2)^2$ and neglecting the minor mixing of $\ket{+}_{\theta},\ket{-}_{\theta}$ with $\ket{0}$, the calculated eigenvectors become: 
\begin{equation}
\label{eq:eigenvectors_comb_B_z}
\begin{split} 
&\ket{-}_{\theta, \mathcal{B}_{\parallel}} = \sin(\frac{\gamma}{2})e^{-i\theta}\ket{S_z=+1} - \cos(\frac{\gamma}{2})e^{i\theta}\ket{S_z=-1} \\
& \ket{+}_{\theta, \mathcal{B}_{\parallel}} = \cos(\frac{\gamma}{2})e^{-i\theta}\ket{S_z=+1} + \sin(\frac{\gamma}{2})e^{i\theta}\ket{S_z=-1}
\end{split}
\end{equation}

While the corresponding eigenergies are: 

\begin{equation}
\label{eq:eigenvalues_comb_B_z}
\begin{split}
& E_{0}^{(2)} = - \frac{\mathcal{B}_{\bot}^2}{2D_{gs}} \\
& E_{-,\theta, \mathcal{B}_{\parallel}}^{(2)} = D_{gs} + \frac{\mathcal{B}_{\bot}^2}{2D_{gs}} - \sqrt{(E_{gap}/2)^2 + \mathcal{B}_{\parallel}^2}\\
& E_{+,\theta, \mathcal{B}_{\parallel}}^{(2)} = D_{gs} + \frac{\mathcal{B}_{\bot}^2}{2D_{gs}} + \sqrt{(E_{gap}/2)^2 + \mathcal{B}_{\parallel}^2} \\
\end{split}
\end{equation}
with $\mathcal{B}_{\parallel} = g\mu_BB_{\parallel}$ and $\tan{\gamma} = \frac{E_{gap}}{2\mathcal{B}_{\parallel}}$ (see Appendix \ref{Calculations eigenvalues} for a derivation of these equations).

For $\mathcal{B}_{\parallel}=0$, one has $\frac{\gamma}{2}=\frac{\pi}{4}$ therefore the eigenstates are balanced superpositions of strong-axial field states. As {$\mathcal{B}_{\parallel}$ increases, $\frac{\gamma}{2}$ decreases, therefore the states become partially-dressed states and acquire a non-zero value of $\langle S_z \rangle$. For larger axial magnetic fields ($\mathcal{B}_{\parallel}^2 \gg (\frac{E_{gap}}{2})^2$ )the eigenergies become linear in $B_{\parallel}$ resulting in strong-axial field states. This behavior can be seen in  Fig. \ref{fig:energytheo}(b). For the moment, we consider only the data regarding nuclear spin $I_z=0$ (red curve), because we are neglecting the effect of the nuclear spin.

Fig. \ref{fig:energytheo}(b) shows the trace distances \cite{nielsen2010quantum}\footnote{The trace distance $D(\rho,\sigma) $ between two density matrices $\rho$ and $\sigma$ is $D(\rho, \sigma) = \frac{1}{2} \text{Tr} \left| \rho - \sigma \right|$ }  between the state $\ket{-}_{\theta}$ and the numerically evaluated eigenstates of the Hamiltonian in Eq. (\ref{eq:tot_ham}). The data regarding nuclear spin $I_z=0$ indicate that the two states are at zero distance for $B_{\parallel}=0$ but, as $B_{\parallel}$ increases, the distance between the states also increases. Additionally, with increasing $B_{\parallel}$ the eigenstates acquire a non-zero value of $\langle S_z \rangle$  (Fig. \ref{fig:energytheo}(c)), as predicted by Eq.s (\ref{eq:eigenvectors_comb_B_z}).
\subsubsection{Complete Hamiltonian}
The interaction between NV electronic spin and $^{14}$N nuclear spin $\vec{I}$ comprises two terms, quadrupolar and hyperfine interactions. The effect of the quadrupolar term is to turn $I_z$ into a good quantum number and to increase the energy of states with $I_z=\pm1$. The effect of the hyperfine term is well known; it shifts the position of the dressed state from ${B_{\parallel}}^{dressed}=0$ to ${B_{\parallel}}^{dressed}=\mp\frac{A_{\parallel}}{g\mu_B}$ for $I_z=\pm 1$. 
This phenomenon enables the simultaneous presence of dressed and partially dressed for the same applied field $B_{\parallel}$.

The analysis of the complete Hamiltonian eigenenergies will focus on two values of $B_{\parallel}$:
\begin{itemize}
	\item $B_{\parallel}=0$ (point A in Fig. \ref{fig:energytheo}). At this point, the eigenstates with $I_z=0$ are dressed states, the eigenstates with $I_z=\pm1$ are degenerate partially-dressed states,  with opposite values of $\langle S_z \rangle$\footnote{to be precise, there is an anticrossing between the $I_z=+1$ and $I_z=-1$ components at $B_{\parallel}=0$. This anticrossing is due to the orthogonal part of the hyperfine term. These states are very fragile. They are destroyed by an axial magnetic field of around 100 nT and the energy gap created by the anticrossing is very small, around 2 kHz.} 
	\item $B_{\parallel}= \frac{A_{\parallel}}{g\mu_B}$ (point B in Fig. \ref{fig:energytheo}). At this point, the eigenstate with $I_z=-1$  is completely dressed, and the two other eigenstates are partially dressed. 
\end{itemize}
As it will be discussed later, working in point A is promising for quantum computation applications, and working in point B is promising for quantum sensing applications.

\section{Experimental results and discussion}

\begin{figure*}[ht!]
\includegraphics[width=\textwidth]{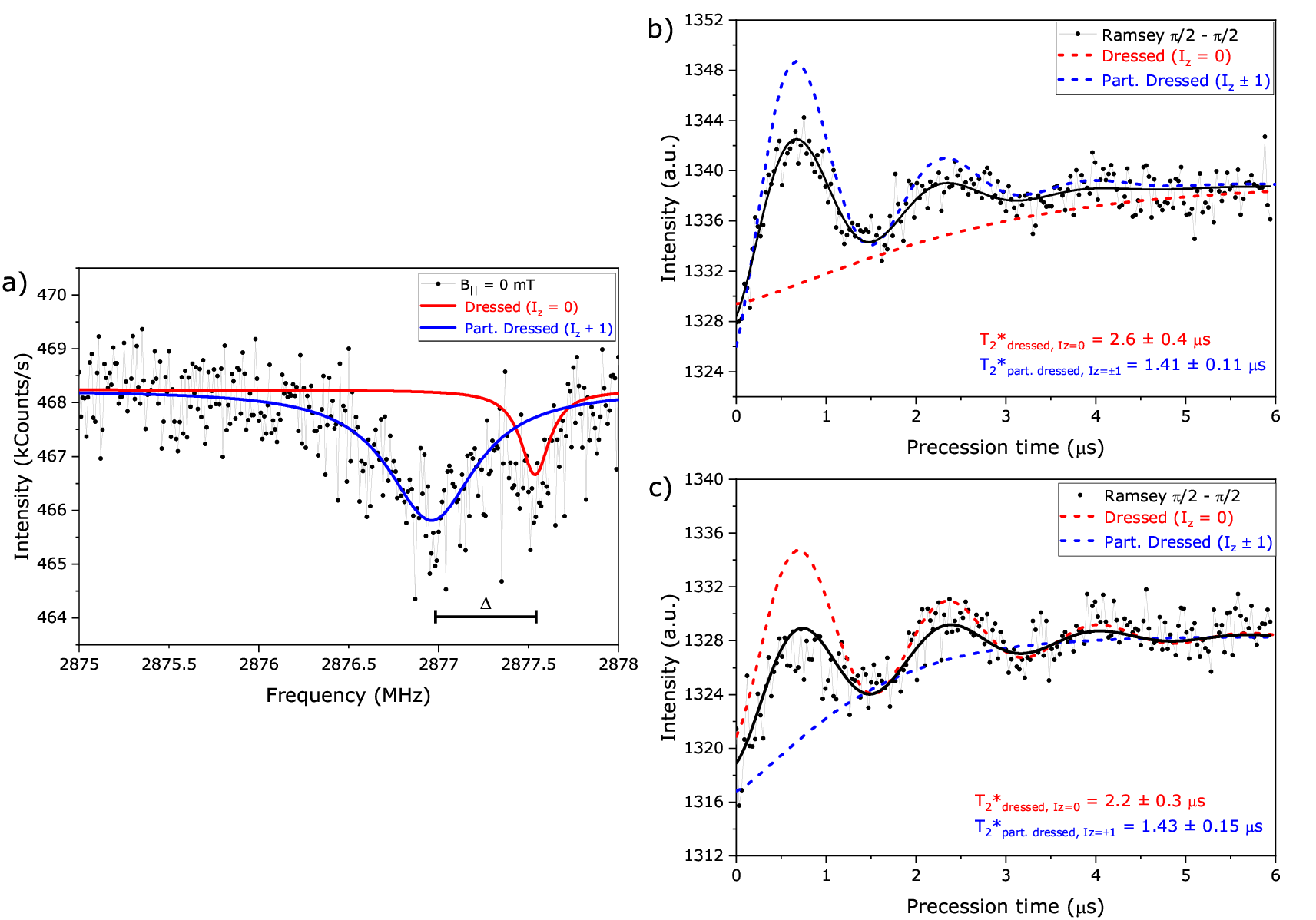}
\caption{Experimental results for $B_{\parallel}=0$, corresponding to point A in Fig. (1): (a) Optically Detected Magnetic Resonance spectra. Two resonances are present. The blue solid line fits the one related to the two degenerate partially-dressed states ($I_z=\pm1$), following a Lorentzian distribution. The red solid line fits the one related to the dressed state ($I_z=0$).  fits this resonance following a Lorentian distribution. (b) Free Induction Decay relaxation for a microwave of frequency $\nu_{MW}=\nu_{A,dres}$ on resonance with the dressed state $I_z=0$. The decay is fitted as the contribution of a pure stretched exponential, red dashed curve, relative to the dressed state $I_z=0$,  and an oscillating stretched exponential, blue dashed curve, related to the partially dressed states with $I_z =\pm 1$. (c) Free Induction Decay relaxation for a microwave of frequency $\nu_{MW}=\nu_{A,p-dres1}$ on resonance with the degenerate partially dressed states $I_z=\pm1$. The decay is fitted as the contribution of a pure stretched exponential, blue dashed curve, related to the partially-dressed states with $I_z=\pm 1$, and an oscillating stretched exponential, red dashed curve, related to the dressed state with $I_z = 0$.
In all curves, the value of the stretched exponential is fixed to $p=1.24$, derived from the fit of the decaying of strong axial field states, see Supplemental Materials.}
\label{fig:expA} 
\end{figure*}
Our experimental investigation was performed on an ensemble of $N \sim 6\cdot10^4$ independent NV centers distributed in a volume $V \sim 1\cdot10^{-17} m^3$, see Supplemental Material. Each NV is subjected to inhomogeneous magnetic and total effective electric fields, described by the distributions $f_{B}(B)$ and $f_{\Pi}( \Pi)$. By reasonably neglecting the interaction between NV centers, the experimental results can be interpreted by solving Eq. (\ref{eq:tot_ham}) for each value of $B$ and $\Pi$. The eigenstate of the ensemble is the tensor product of the single NV density matrices, with each density matrix weighted according to $P(B)$ and $P(\Pi)$. The eigenenergies of the ensemble are the weighted sum of a single NV eigenergies according to $f_{B}(B)$ and $f_{\Pi}( \Pi)$. The resonant frequency $\nu_{res}$, which is computed from the differences in the eigenergies, also follows a distribution $f_{\nu_{res}}(\nu_{res})$. In the following, two sets of results are presented: the first is related to CW microwave excitation, and the second to pulsed measurements (i.e. FID).

\subsubsection{Cw excitation}
In a CW-ODMR measurement (see Fig.\ref{fig:expA}(a) and \ref{fig:expB}(a)), an oscillating magnetic field with frequency $\nu_{MW}$ and a 532 nm laser are applied simultaneously, exploiting a conventional single-photon confocal microscope with the light detected by a single photon detector (see \cite{Babashah2023rev} and Supplemental Materials). Materials for details on the experimental setup). The laser excitation has two effects: (i) inducing red photoluminescence (PL), whose intensity depends on the spin state of the NV, with $\ket{0}$ being brighter than $\ket{-}_{\theta}$ and $\ket{+}_{\theta}$; and (ii) polarizing the NV in spin state $\ket{0}$. To explain the technique, let's focus on states $\ket{0}$ and $\ket{-}_{\theta}$. If $\nu_{MW}$ is far from the resonant frequency $\nu_{res}=\left(E_{\ket{0}} - E_{\ket{-}_{\theta}}\right)/\hbar$, the oscillating magnetic field is not effective in driving transitions between $\ket{0}$ and $\ket{-}_{\theta}$, moreover, the green light polarizes the NV state to $\ket{0}$, resulting in brighter PL emission. Conversely, when $\nu_{MW}=\nu_{res}=\left(E_{\ket{0}} - E_{\ket{-}_{\theta}}\right)/\hbar$, the oscillating field is very effective in driving transitions between $\ket{0}$ and $\ket{-}_{\theta}$, therefore the dark state $\ket{-}_{\theta}$ is populated and the collected PL presents a minimum \cite{dreau2011avoiding}.

In Fig. 1 (point A), two resonances are observable for $B_{\parallel}=0$, one related to a dressed state, $\nu_{A,dres}$, and the other, $\nu_{A,p-dres1}$, related to two degenerate partially-dressed states with opposite values of $\langle S_z \rangle$, as predicted numerically (see Fig. \ref{fig:energytheo} (c) and (d)). The two transitions are clearly visible in the ODMR spectrum in Fig. \ref{fig:expA}(a), the difference $\Delta=\nu_{A,dres}-\nu_{A,p-dres1}$ between the two resonance frequencies is indicated. The resonance corresponding to $I_z=\pm1$ presents a greater contrast due to the presence of the two degenerate populations. In point B, for $B_{\parallel}=\frac{A_{\parallel}}{g\mu_B}$, there are three resonant frequencies: one corresponding to a dressed state, $\nu_{B,dres}$, and two corresponding to partially dressed states with different non-zero values of $\langle S_z \rangle$, $\nu_{B,p-dres2}\,(I_z=0)$ and $\nu_{B,p-dres3}\,(I_z=+1)$. The three resonances are clearly visible in the ODMR spectrum in Fig. \ref{fig:expB} (a). In this case, the resonances present similar contrasts because each resonance corresponds to a single state. The three frequency differences $\Delta_1=\nu_{B,dres}-\nu_{B,p-dres2}$, $\Delta_2=\nu_{B,p-dres2}-\nu_{B,p-dres3}$, $\Delta_3 =\nu_{B,dres}-\nu_{B,p-dres3}$ are indicated. It is important to underline that $\nu_{A,dres}=\nu_{B,dres}$ because in point B the axial field compensates the hyperfine term, while the contribution of the nuclear quadrupolar term is equivalent both for $\ket{0,1}$ and $\ket{-_{\theta},1}$. Therefore it does not enter in the derivation of $\nu_{B,dres}$.

The different states are excited by different polarizations of the oscillating field. This is interesting because in an NV  ensemble where dressed, partially dressed, and high field states are present, the different polarizations enable to be selective on states excitation. Considering Rabi and FID measurements, dressed states are excited by an oscillating magnetic field which is linearly polarized in the x-y plane and aligned along the direction defined by the angle $\theta$ in Eq \ref{eq:theta}. Strong axial field states are excited by an oscillating magnetic field that is circularly polarized in the x-y plane, and that can be viewed as the balanced sum of two oscillating fields linearly polarized along orthogonal directions. Partially dressed states are polarized by an oscillating magnetic field elliptically polarized in the x-y plane, which is as the unbalanced sum of two oscillating fields, one polarized along $\theta$ and the other along a direction orthogonal to $\theta$, see Appendix \ref{Calculation Polarization} for further details. An oscillating magnetic field linearly polarized in the x-y plane along an angle different from $\theta$ is used. This oscillating magnetic field can be viewed as the sum of two linearly polarized fields, one polarized along $\theta$ and the other in a direction orthogonal to $\theta$. In this way, we can excite both dressed and partially dressed states.

CW-ODMR spectra hint at the different behaviors of dressed and partially dressed states. Notably, the former ones exhibit narrower transition linewidths with respect to the latter ones. This is because the dressed states are sensitive only at the second order to noise coming from coupling with the spin baths and spatial inhomogeneities and temporal fluctuations in the external magnetic field $B$. When these are the two main sources of decoherence, dressed states are expected to show longer coherence times $T_2^*$ (see Appendix \ref{Calculation dec single}, \ref{Calculation dec ensemble}). 
CW-ODMR measurements are not ideal for discriminating between dressed and partially dressed states due to the effect of power broadening: to clearly resolve peaks linked to the different states, it is necessary to decrease the microwave (MW) power. However, this decreases the contrast and reduce the overall S/N ratio. Hence, Free Induction Decay(FID) measurements are selected to characterize the different types of states, since these are not affected by MW-power broadening and can give a direct evaluation of $T_2^*$.

\subsubsection{Pulsed measurements}
In general, FID measurement procedures consist in an initialization optical pulse, two $\frac{\pi}{2}$ MW pulses separated by a free precession interval of duration $\tau$ and a final readout optical pulse (see Supplementary Material). The system is initialized in the state $\ket{0}$, the first $\frac{\pi}{2}$ pulse brings the state in a superposition of $\ket{0}$ and $\ket{S_z \neq 0}$. During the free evolution time, the two components of the superposition acquire a phase difference $\phi$ depending on the detuning $\Delta=\nu_{MW}-\nu_{res}$ between MW excitation frequency $\nu_{MW}$ and the resonant frequency $\nu_{res}$. The second $\frac{\pi}{2}$ encodes $\phi$ in the population of the $\ket{0}$ state ($p_{\ket{0}}$)\cite{Bifone2018relaxometry}. The final read-out optical pulse excites a PL proportional to $p_{\ket{0}}$. 
The general form of  $p_{\ket{0}}$ for a FID measurement where a single resonance is driven is: 

\begin{equation} 
p_{\ket{0}}(\tau)=\frac{1}{2}\left[1-e^{-\left(\frac{\tau}{T_2^*}\right)^p}\cos(2\pi\Delta\tau)\right]
\label{eq:PL}
\end{equation}
 
with $1\leq p<2$. The coherence time $T_2^*$ defines the timescale at which the system loses quantum coherence, and $p_{\ket{0}}$ decays to $\frac{1}{2}$, corresponding to a completely mixed state \footnote{In the usual terminology $T_2^*$ is the time-constant that describes the loss of coherence considering field inhomogeneities. $T_2$ is the time constant that describes the loss of coherence due solely to intrinsic sources. $T_1$ is the time constant related to the relaxation to thermal equilibrium.}.
For an ensemble of NVs, the decay is caused by different sources of decoherence: coupling with surrounding spins, temporal fluctuations and spatial gradients in the external fields $B_i$ and $\Pi_i$ \cite{bauch2020decoherence, dobrovitski2008decoherence}. The value of $p$ provides information on the relative weight of the different sources of decoherence: if $p\sim1$ the main source of decoherence is coupling with surrounding spins, while an increase in $p$ corresponds to a greater influence of temporal fluctuations and spatial gradients of the external fields \cite{bauch2018ultralong,bauch2020decoherence,dobrovitski2008decoherence}. If the microwave pulse $\frac{\pi}{2}$ has enough spectral width \footnote{The spectral width of $\frac{\pi}{2}$ pulse is approximately equal to the Rabi frequency of the MW} to drive more than one resonance, the PL signal is the sum of multiple terms analogous to Eq.(\ref{eq:PL}) but with different $\Delta_i=\nu_{MW}-\nu_{res,i}$.

\begin{figure*}[t!]
\includegraphics[width=\textwidth]{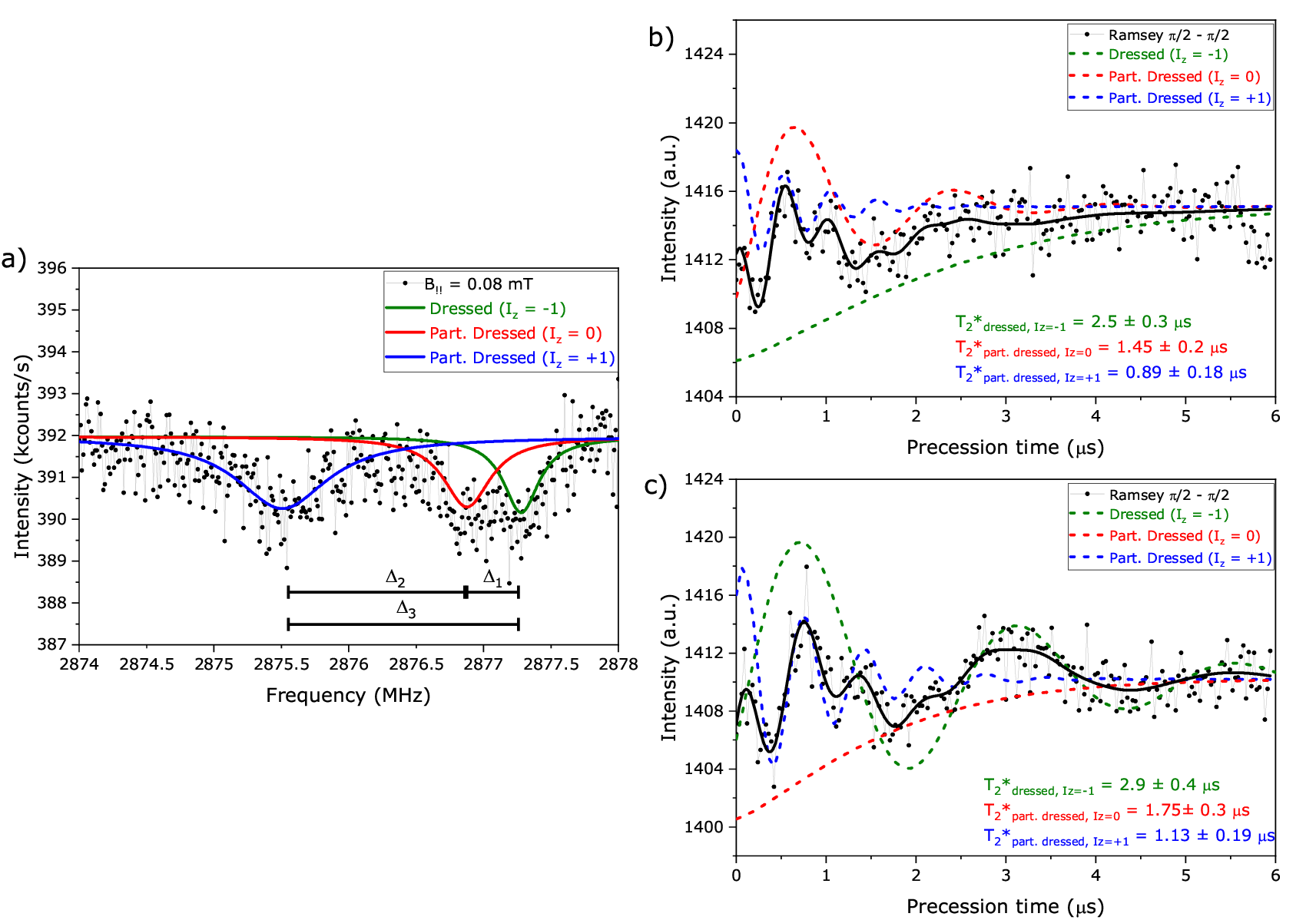}
\caption{Experimental results for $B_{\parallel}=0.8 $ mT, corresponding to point B in Fig. (1):  (a) Optically Detected Magnetic Resonance spectrum. Three resonances are present. The blue solid line fits the resonance related to a partially-dressed state ($I_z = +1 $). The red solid line fits the resonance related to another partially-dressed state ($I_z = 0 $). The one at the larger frequency $\nu_{B,dres}$ is related to the dressed state ($I_z = -1$). The green solid line fits this resonance following a Lorentzian distribution. 
(b) Free Induction Decay relaxation for a microwave of frequency $\nu_{MW}=\nu_{B,dres}$ on resonance with the dressed state $I_z=-1$. The decay is fitted as the contribution of a very slow oscillating stretched exponential, green dashed curve, and two faster oscillating stretched exponentials, red dashed curve and blue dashed curve. The very slow stretched exponential is related to the dressed state with $I_z=-1$. The intermediate oscillating exponential is related to the partially dressed 
state with $I_z=0$. The faster oscillating exponential is related to the partially dressed state with $I_z=+1$. 
(c) Free Induction Decay relaxation for a microwave of frequency $\nu_{MW}=\nu_{B,p-dres2}$ on resonance with the dressed state $I_z=0$. The decay is fitted as the contribution of a pure stretched exponential, red dashed curve, and two oscillating stretched exponentials, green dashed curve and blue dashed curve.
The pure stretched exponential is related to the partially dressed state with $I_z = 0$.
The slower oscillating exponential is related to the dressed state with $I_z=-1$, and the faster oscillating exponential is related to the partially dressed state with $I_z=+1$.The value of the stretched exponential is fixed to $p=1.24$, derived from the fit of the decaying of strong axial field states}
\label{fig:expB} 
\end{figure*}

The first set of pulsed measurements (Fig. \ref{fig:expA}) was collected with $B_{\parallel}=0$, corresponding to point A in Fig. \ref{fig:energytheo}. The spectral width is sufficient to excite both resonances. First, we consider a FID measurement on resonance with the dressed state with $I_z=0$, $\nu_{MW}=\nu_{A,dres}$, shown in Fig. \ref{fig:expA} (b). The experimental curve is the sum of a pure stretched exponential and an oscillating stretched exponential. The pure stretched exponential is associated with a detuning $\Delta=0$, corresponding to the resonance of the dressed state. The explicit expression for the decay function used in the fit is 
\begin{equation} 
f(y)=y_0 + Ae^{\left(-t/T_{2,\text{dressed}\,\, I_z=0}^*\right)^p}. 
\end{equation}
The oscillating exponential is associated with a detuning $\Delta=532$ kHz, corresponding to the resonance of a partially-dressed state. The explicit expression for the decay function used in the fit is 
\begin{equation} 
f(y)=y_0 + Ae^{\left(-t/T_{2,\text{part. dressed-1}\,\, I_z= \pm 1 }^*\right)^p}\cos(2 \pi \Delta t + \phi). 
\end{equation}
The estimated coherence time of dressed states is 1.3 times longer than that of the partially dressed state, as listed in Table \ref{tab:coherencetime}. 

A similar improvement in $T_2^*$ is observed when the FID measurement is performed on resonance with the degenerate partially-dressed states with $I_z=\pm1$, $\nu_{MW}=\nu_{A,p-dres1}$, as shown in Fig. \ref{fig:expA} (c). In this case, the pure exponential is related to the partially-dressed state. The explicit expression for the decay function used in the fit is 
\begin{equation} 
f(y)=y_0 + Ae^{\left(-t/T_{2,\text{part. dressed-1}\,\, I_z=\pm 1}^*\right)^p}
\end{equation}
The oscillating exponential is related to the dressed state. The explicit expression for the decay function used in the fit is 
\begin{equation} 
f(y)=y_0 + Ae^{\left(-t/T_{2,\text{dressed}\,\, I_z= \pm 1 }^*\right)^p}\cos(2 \pi \Delta t + \phi)
\end{equation}

Moreover, the values of $T_2^*$ for dressed states are compatible among measurements performed on resonance with different transitions. This behavior can be observed also for the values of $T_2^*$ for partially-dressed states. A quick observation of the experimental decays in Fig. \ref{fig:expA} confirms the previous finding: oscillations last longer when the detuned state is the dressed state. 

FID measurements were also conducted for another sub-ensemble with a different NV-axis orientation, as detailed in the Appendix \ref{Calculations High Field}, that exhibited a component along the z-axis $B_{\parallel} \approx 3$ mT. The FID data were recorded by tuning the MW frequency on resonance with the central hyperfine peak, thus only a single detuning ($\nu=2.16$ MHz) is observed, see Fig. \ref{fig:FID_High_Field}. As described in the previous section, strong axial field states couple with the spin bath and fluctuations of external magnetic fields, leading to shorter coherence times than dressed states. In this case, the axial field is much larger than the orthogonal field, therefore the eigenstates are pure strong-axial-field states, resulting in $T_2^*$ that is more than two times shorter than the one obtained for dressed states, as shown in Table \ref{tab:coherencetime}.

Dressed states, partially-dressed states and strong-axial field states present different coherence times as they are affected differently by the various sources of decoherence. Dressed states are more sensitive to temporal fluctuation and inhomogeneities of the total electric field $\vec{\Pi}$, because $\langle S_{x}^2-S_{y}^2\rangle\neq0, \,\, \langle S_xS_y + S_yS_x \rangle \neq 0$ which are the coupling terms for $\Pi_x$ and $\Pi_y$ in the Hamiltonian presented in Eq. (\ref{eq:tot_ham}), while $\langle S_z \rangle=0$. Strong-axial field states are more sensitive to dipolar coupling with the surrounding spin bath, temporal fluctuation and spatial gradients of the axial magnetic field $B_{\parallel}$ because $\langle S_z \rangle \neq 0$ while $\langle S_{x}^2-S_{y}^2\rangle, \,\, \langle S_xS_y + S_yS_x \rangle  = 0$. On the other hand, for partially dressed state, $\langle S_z \rangle \neq 0$ and $\langle S_{x}^2-S_{y}^2\rangle, \,\, \langle S_xS_y + S_yS_x \rangle \neq 0$, therefore the different sources of decoherence compete. For more details on this topic, see \cite{jamonneau2016competition} and Appendix \ref{Calculation dec ensemble}. The decrease in coherence time  $T_2^*$ observed when passing from dressed states to partially dressed states then to strong axial field states indicates that dipolar coupling with surrounding spin baths and temporal fluctuations and spatial inhomogeneities of the axial magnetic field are the major sources of decoherence in the sample being studied. This can be explained by the high concentration of $^{14}N$ spin centers that couple to NVs as a consequence of the large implantation fluence employed ($F=1\cdot10^{14}$ cm$^{-2}$).
The stretched exponential parameter $p$ should differ for dressed, partially dressed, and strong axial field states. Dressed states are more influenced by inhomogeneities and 
temporal fluctuations of $\vec{\Pi}$; therefore, a value of $p$ close to 2 is expected. For strong axial field states, instead, a value of $p$ close to 1 is 
expected due to the effect of the spin bath. In general, the value of $p$ is set by the competition between these decoherence sources and the values of field gradients. 
For simplicity, the value $p=1.28$ obtained from the single decay of strong-axial field states is used as a fixed parameter for decays of both dressed and partially dressed states, taking into account that a considerable uncertainty is associated with 
the estimation of $p$ (see Supplemental Materials).

\begin{table*}[t]
\centering
\begin{tabular}{|c|c|c|c|c|c|}                                              
\hline                      
                                                                                                                                  & $T_2^*$ dressed & $T_2^*$ part-dressed-1 &  $T_2^*$ part-dressed-2 &  $T_2^*$ part-dressed-3 & $T_2^*$ Strong-axial Field\\
& [$\mu$s] & [$\mu$s] & [$\mu$s] & [$\mu$s] & [$\mu$s] \\
\hline
$B_{\parallel}=0 \text{(A)},\,\,\nu_{MW}=\nu_{A,dres}$                                                           & $2.6\pm0.4$       & $1.41\pm0.11$               &  -                                      &  -                                    & -                            \\
$B_{\parallel}=0 \text{(A)},\,\,   \nu_{MW}=\nu_{A,p-dres1}$                                                   & $2.2\pm0.3$       & $1.43\pm0.15$              &  -                                       & -                                     & -                            \\ 
$B_{\parallel} =\frac{A_{\parallel}}{g\mu_B} \text{(B)} ,\,\,   \nu_{MW}=\nu_{B,dres}$                                                       & $2.3\pm0.3$       & -                                     &  $1.43\pm0.15$                &  $0.89\pm0.18$             & -                            \\
$B_{\parallel} = \frac{A_{\parallel}}{g\mu_B} \text{(B)} ,\,\,  \nu_{MW}=\nu_{B,p-dres2}$       & $2.9\pm0.4$       & -                                     &  $1.75\pm0.15$                &  $1.13\pm0.19$             & -                            \\ 
Strong-axial Field                                                                                                       & -                          & -                                     &  -                                       &  -                                    & $0.88.\pm0.17$      \\ 
\hline
\end{tabular}
\caption{Values of coherence time $T_2^*$, derived from the fit of the Free Induction Decays for different experimental conditions. Each row corresponds to a value of axial field $B_{\parallel}$ and of microwave frequency $\nu_{MW}$.}
\label{tab:coherencetime}
\end{table*}
The main result of this work is described by the experimental data shown in Fig.\ref{fig:expA}, which clearly indicate the ability to coherently drive two resonance states associated with different values of the nuclear spin $I_z$, while extending the coherence time $T^*_2$ with respect to the typically used strong-axial field state. This opens up the potential for leveraging the weak orthogonal field regime to achieve high fidelity C$_n$NOT$_e$ gates, which are essential for quantum computation and have also been recently employed to enhance the sensitivity of magnetic measurements \cite{arunkumar2023quantum}. A basic C$_n$NOT$_e$ gate can be realized by employing the FID sequence on resonance with the dressed state with $I_z=0$, and adjusting the free evolution time to $\tau=\frac{\pi}{\Delta}$, which corresponds to the half period of the decaying oscillations. 

In more detail, if we start from the state $\ket{S_z=0}\otimes\ket{I_z=0}$, at the end of the FID sequence, we will have the state  $\ket{S_z=-}\otimes \ket{I_z=0}$. We are on resonance, and so the superposition $ \left(\ket{S_z=0}+\ket{S_z=-}\right)\otimes \ket{I_z=0}$, created by the first $\frac{\pi}{2}$, does not precede, but it is only subjected to dephasing. The second $\frac{\pi}{2}$ pulse create the state $\ket{S_z=-}\otimes \ket{I_z=0}$, if we neglect the dephasing. The FID sequence acts as a $\pi$ pulse. Instead, if we start from the state $\ket{S_z=0}\otimes\left(\ket{I_z=+1}+\ket{I_z=-1}\right)$, the first $\frac{\pi}{2}$ pulse create the superposition $\left(\ket{S_z=0}+\ket{S_z=-_{\theta,B_{\parallel}}}\right)\otimes \left(\ket{I_z=+1}+\ket{I_z=-1}\right)$. This superposition precedes because we are not in resonance. After a time $\tau=\frac{\pi}{\Delta}$ we have the state $\left(\ket{S_z=0}-\ket{S_z=-_{\theta,B_{\parallel}}}\right)\otimes \left(\ket{I_z=+1}+\ket{I_z=-1}\right)$. The second $\frac{\pi}{2}$ pulse recover the initial state $\ket{S_z=0}\otimes\left(\ket{I_z=+1}+\ket{I_z=-1}\right)$. The FID sequence acts as an identity.  

We remind that dressed states are sensitive to the magnetic field only at the second order, while, the partially-dressed states are sensitive to the magnetic field at the first order, being characterized by a non-zero expectation value of the axial spin $\langle S_z \rangle\neq0$. This paves the way for applications in magnetic field sensing. For this kind of application, working in point A in Fig. \ref{fig:energytheo} is not ideal, since, the degenerate partially-dressed states correspond to two opposite values of  $\langle S_z\rangle$, and a magnetic field $B^{sense}_{\parallel}$ will cause two opposite detunings \footnote{we are, again, neglecting the effect of the anticrossing.}. The effect of these detunings will tend to cancel each other in a typical magnetic field measurement. 

For these reasons, we apply a small bias magnetic field $B_{\parallel}=\frac{A_{\parallel}}{g\mu_B}=0.08$ mT, corresponding to point B of Fig. \ref{fig:energytheo}, where there are three resonances, one $\nu_1$, belonging to a dressed state, and two $\nu_2$ and $\nu_3$ belonging to two partially-dressed states with two different values of $\langle S_z\rangle$. The results of FID measurement performed under this regime are shown in Fig. \ref{fig:expB}(b)-(c). Fig.\ref{fig:expB}(b) shows a measurement taken on resonance with the $I_z=-1$ state ($\nu_{mw}=\nu_{B,dres}$), where a combination of three resonance contributions properly describes the experimental data. One component is a very slow oscillating exponential, associated with the on-resonance dressed state with $I_z=-1$. This slow oscillation of frequency $\delta_1=94$ kHz is due to non-optimal experimental conditions. The explicit expression of the decay function used in the fit is 
\begin{equation}
f(y)=y_0 + Ae^{\left(-t/T_{2,\text{dressed}\,\, I_z=-1}^*\right)^p}\cos(2 \pi \delta_1 t + \phi)
\end{equation}

Then, there are two faster-oscillating exponentials corresponding to detunings $\Delta_1=\nu_{B,dres}-\nu_{B,p-dres2}$ and $\Delta_3=\nu_{B,dres}-\nu_{B,p-dres3}$, which belong to two partially dressed states. The expressions of the decays functions used in the fit are:
\begin{equation}
f(y)=y_0 + Ae^{\left(-t/T_{2,\text{part. dressed-2}\,\, I_z=0}^*\right)^p}\cos(2 \pi \Delta_1 t + \phi)
\end{equation}
for the partially dressed state with $I_z=0$, and 
\begin{equation}
f(y)=y_0 + Ae^{\left(-t/T_{2,\text{part. dressed-3}\,\, I_z=0}^*\right)^p}\cos(2 \pi \Delta_3 t + \phi)
\end{equation}
for the partially dressed state with $I_z=+1$.
The different components exhibit different $T_2^*$ values. Similarly to point A, the dressed states exhibit a longer $T_2^*$ than partially dressed states. Moreover, $T_2^*$ decreases moving from partially-dressed state 2 to partially-dressed state 3 because of an associated increase in $\langle S_z\rangle$, see Table \ref{tab:coherencetime}. 

A similar behavior in $T_2^*$ is observed when FID measurements are performed on resonance with partially-dressed state having $I_z= 0$ ($\nu_{MW}=\nu_{B,p-dres2}$), see Figure \ref{fig:expB}(c). In this case, there is one very slow oscillating exponential associated with the on-resonance partially-dressed state $I_z=0$, at a frequency $\delta_2=147$ kHz. Again, this slow oscillation is due to non-optimal experimental conditions. The  expression of the decay used in the fit is 
\begin{equation}
f(y)=y_0 + Ae^{\left(-t/T_{2,\text{part. dressed-2}\,\, I_z= 0}^*\right)^p}\cos(2 \pi \Delta_2 t + \phi)
\end{equation}
Then there are two faster oscillating exponential corresponding to the detunings $\Delta_1=\nu_{B,p-dres2}-\nu_{B,dres}$ and $\Delta_2=\nu_{B,p-dres2}-\nu_{B,p-dres3}$. The expressions of the decays used in fit are 
\begin{equation}
f(y)=y_0 + Ae^{\left(-t/T_{2,\text{dressed}\,\, I_z=-1}^*\right)^p}\cos(2 \pi \Delta_1 t + \phi)
\end{equation}
for the partially dressed state with $I_z=-1$ and  
\begin{equation}
f(y)=y_0 + Ae^{\left(-t/T_{2,\text{part. dressed-3}\,\, I_z=0}^*\right)^p}\cos(2 \pi \Delta_2 t + \phi)
\end{equation}
for the partially dressed state with $I_z=0$. Under these experimental conditions as well, the value of $T_2^*$ depends on the degree to which a state is dressed and not on the value of the driving microwave frequency $\nu_{MW}$.

We suggest that a sensing regime working in point B in Fig. 1 represents an interesting alternative to current methods\cite{toyli2013fluorescence,neumann2013high,fang2013high,kehayias2014microwave} to decouple the effects of different external variables (e.g. magnetic, electric field, strain, temperature, etc) on the ODMR resonance. Using quantum optimal control techniques \cite{oshnik2022robust} it is possible to selectively drive the magnetic-independent resonance $\nu_{B,dres}$ or the magnetic dependent resonance $\nu_{B,p-dres2}$. For instance, to detect a variation of axial magnetic field $\Delta B_{\parallel}$ in the presence of a temperature variation $\Delta T$, it is possible to measure the detuning $\Delta_1$, which is not sensitive to $\Delta B_{\parallel}$ and use it as a reference to subtract detuning $\Delta_2$ that depends on $\Delta B_{\parallel}$. The two resonances can be driven selectively by using a single microwave frequency modulated pulse with a bandwidth of 5 MHz, as they are separated only by $\nu_{B,dres} - \nu_{B,p-dres2}=532$ KHz. Another possibility is to create a superposition of nuclear states, $\ket{S_z=0}\otimes\frac{1}{2}(\ket{I_z=+1}+\ket{I_z= 0})$, and then apply a sequence of rectangular pulses that excite both resonances, similar to the ones used for quantum beats magnetometry and thermal echo. Generally, with these techniques, the states being simultaneously affected are the typical strong-axial field states $\ket{S_z= +1}$ and $\ket{S_z=-1}$, while in this case, the dressed and partially dressed states would be used. The primary advantage of our proposal is that the dressed and partially-dressed states are, at working point B, closer in frequency compared to the $\ket{S_z= +1}$ and $\ket{S_z=-1}$ states in strong axial field configuration. This allows for lower power pulses, providing an advantage, for instance, in sensing biological systems.

\section{Conclusions}

Free Induction Decay (FID) measurements have been carried on in the presence of a weak orthogonal field and a total electric field for an ensemble of Nitrogen-Vacancy centers. First, the competition between the weak orthogonal field and the total electric field was considered, showing that the resulting eigenstates are dressed states, balanced superposition of strong axial field states, rotated in the orthogonal plane. The explicit formula for the rotation angle was obtained from the expression of the exact eigenstates (up to now in literature only calculations in the perturbative approach were available). Then, the role of axial magnetic fields in creating unbalanced superposition of strong axial field states, which we call partially-dressed states, was described. Two working points were studied experimentally: one with a null axial magnetic field applied and a second with an axial magnetic field matching the hyperfine field caused by the $^{14}N$ nucleus. In both working points, we observed the presence of dressed and partially dressed states in the FID measurements. Dressed and partially dressed states can be distinguished by different coherence time $T_2^*$, which shows a decrease when transitioning from dressed to partially dressed states. Compared to the widely studied strong-axial field states, the coherence time $T_2^*$ is enhanced in dressed and partially dressed states. The possibility to simultaneously drive dressed and partially dressed states using a single microwave opens up interesting applications in quantum computation and quantum sensing, see e.g Ref.\cite{Walsworth2023readout} where a repetitive readout protocol on NV ensemble is exploited. We can use the findings of the present paper for the selective magnetic field sensing part and the C$_n$NOT$_e$ previously described to write the state on the N nucleus. 

\begin{acknowledgments}
E.B. thanks A. Rigamonti for helpful discussions. This research has been carried on in the context of 23NRM04 NoQTeS (European Partnership on Metrology  project, co-financed from the European Union’s Horizon Europe Research and Innovation Programme and by the Participating States); Projects Qutenoise of Trapezio of San Paolo Foundation; Experiment QUISS funded by INFN CSN5. This project has received funding from the European Union’s Horizon Europe – The EU research \& innovation programme under the Grant Agreement number 101189611. This research has been funded by the Italian Ministry of University and Research (MUR), “NEXT- GENERATION METROLOGY”, FOE 2023 (Ministry Decree n. 789/2023). The authors wish to acknowledge the National Collaborative Research Infrastructure Strategy (NCRIS) funding provided by the Australian Government for this research. J.F. acknowledges support from MUR project PE \_00000023 -NQSTI - SPOKE 4- CUP B53C22004170006. The authors also wish to thank F. Saccomandi for the technical support in developing the magnetic field control system.
\end{acknowledgments}


\appendix
\clearpage
\section{Competition between Orthogonal Total Electric Field and Orthogonal Magnetic Field (Exact Solution)}
\label{Calculations eigenvalues}
\subsection{Problem Set-up}
We consider the Hamiltonian:
\begin{widetext}
\begin{equation}
	\label{eq:ham_elec}
	\mathcal{H}_{elec-orth} = D_{gs} - d_{\bot}\left[\Pi_x(S_x^2- S_y^2)-\Pi_y(S_xS_y + S_yS_x)\right] \\ +g_e\mu_BB_xS_x  + g\mu_BB_yS_y
\end{equation}
\end{widetext}
where $\vec{B}$ is the magnetic field and  $\vec{\Pi}=\vec{E} + \vec{\sigma}$ is the total electric field that encompasses both the effect of the static electric field and of the strain.
This Hamiltonian in the basis $\ket{S_z=+1}$, $\ket{S_z=0}$, $\ket{S_z=-1}$ can be rewritten as:
\begin{widetext}
\begin{align}
		H_{elec-orth} = &
	\begin{pmatrix}
		D_{gs} & \frac{\mathcal{B}_{\bot}[\cos(\phi_{B})-i\sin(\phi_{B})]}{\sqrt{2}} & -\mathcal{E}[\cos(\phi_{\Pi}) + i \sin(\phi_{\Pi})]\\
		\frac{\mathcal{B}_{\bot}[\cos(\phi_{B})+i\sin(\phi_{B})]}{\sqrt{2}} & 0 & \frac{\mathcal{B}_{\bot}[\cos(\phi_{B})-i\sin(\phi_{B})]}{\sqrt{2}}\\
		-\mathcal{E}[\cos(\phi_{\Pi}) - i \sin(\phi_{\Pi})] & \frac{\mathcal{B}_{\bot}[\cos(\phi_{B})+i\sin(\phi_{B})]}{\sqrt{2}} & D_{gs}\\
	\end{pmatrix}
\end{align}
\end{widetext}
where we used the quantities defined in Eq.s (\ref{eq:parameters}) and (\ref{eq:angles}). \\
\subsection{Computation of eigenvalues}
The eigenvalues of the Hamiltonian in Eq. (\ref{eq:ham_elec}) are obtained by solving:
\begin{equation}
\label{eq:eigenvalues_exact}
	\det (H_{elec-orth}-\lambda\mathbb{I})=0
\end{equation}
that leads to a cubic equation in $\lambda$. The three exact solutions correspond to the three exact eigenvalues $E_{0,exact}, E_{1,exact}, E_{2,exact}$.\\ We have rewritten each eigenvalue as a series expansion of $\zeta=\frac{\mathcal{B}_{\bot}}{D} \ll 1$ where we exploited the ancillary parameter $R= \frac{\mathcal{E}}{\frac{\mathcal{B}_{\bot}^{2}}{D}}=\frac{\mathcal{E}}{\zeta^{2}D} $ to describe the competing effect between $\mathcal{B}_{\bot}$ and $\mathcal{E}$ in the physics of the NV center. Then, only the terms till the second order in $\zeta$ were kept. The resulting approximated eigenvalues are:
\begin{widetext}
\begin{equation}
\label{eq:eigenvalues_approx}
\begin{split}
 E_0 &= -D_{gs}\zeta^2= - \frac{\mathcal{B}_{\bot}^2}{D_{gs}}\\
 E_1 &= D_{gs} + \frac{1}{2}D_{gs}\zeta^2-\left[R^2 + \frac{1}{4} - R \cos(2 \phi _{B_{\bot}} +\phi_{\Pi}) \right]^\frac{1}{2} D_{gs}\zeta^2= D_{gs} + \frac{\mathcal{B}_{\bot}^2}{2D_{gs}}-\left[ \left(\frac{\mathcal{B}_{\bot}^2 }{2D_{gs}}\right)^2 + \mathcal{E}^2 -  \mathcal{E}\frac{\mathcal{B}_{\bot}^2 }{D_{gs}}\cos(2 \phi _{B_{\bot}} +\phi_{\Pi}) \right]^\frac{1}{2}\\
E_2 &= D_{gs} + \frac{1}{2}D_{gs}\zeta^2+\left[R^2 + \frac{1}{4} - R \cos(2 \phi _{B_{\bot}} +\phi_{\Pi}) \right]^\frac{1}{2} D_{gs}\zeta^2=D_{gs} + \frac{\mathcal{B}_{\bot}^2}{2D_{gs}}+\left[ \left(\frac{\mathcal{B}_{\bot}^2 }{2D_{gs}}\right)^2 + \mathcal{E}^2 -  \mathcal{E}\frac{\mathcal{B}_{\bot}^2 }{D_{gs}}\cos(2 \phi _{B_{\bot}} +\phi_{\Pi}) \right]^\frac{1}{2}\\ 
\end{split}
\end{equation}
\end{widetext}
These eigenvalues are the same as the one presented in ref.\cite{doherty2012theory}, a part a factor $\frac{1}{2}$ in first eigenvalue $E_0$.\\ We underline that the parameter $R$ quantifies the relative strength of $\mathcal{B}_{\bot}$ and $\mathcal{E}$. The choice of $R$ expression was inspired by the fact that the term out of diagonal, like $\mathcal{B}_{\bot}$, appears at second-order in standard perturbation theory.
\clearpage
\subsection{Computation of eigenstates}
Starting from the exact eigenvalues of energy, obtained by solving Eq. (\ref{eq:eigenvalues_exact}), we obtained the exact eigenvectors. Here we present their first order in $\zeta$:
\begin{widetext}
\begin{equation}
\label{eq:eigenvectors_approx}
\begin{split} 
 \ket{0} &= \left(\frac{\mathcal{B}_{\bot}}{D_{gs}},-\sqrt{2}e^{i \phi _{B_{\bot}}}, e^{i 2 \phi _{B_{\bot}}}\frac{\mathcal{B}_{\bot}}{D_{gs}}\right) \\\\
 \ket{1} &= \left( 1, \frac{1}{\sqrt2}e^{i \phi _{B_{\bot}}}\left(1-\frac{1-2Re^{i(2 \phi _{B_{\bot}} +\phi_{\Pi})}}{\left| 1-2Re^{i(2 \phi _{B_{\bot}} +\phi_{\Pi})}\right|^2}\right)\frac{\mathcal{B}_{\bot}}{D_{gs}}, -e^{i 2 \phi _{B_{\bot}}}\frac{1-2Re^{-i(2 \phi _{B_{\bot}} +\phi_{\Pi})}}{\left| 1-2Re^{-i(2 \phi _{B_{\bot}} +\phi_{\Pi})}\right|^2} \right) \\
 \ket{2} &= \left( 1, \frac{1}{\sqrt2}e^{i \phi _{B_{\bot}}}\left(1+\frac{1-2Re^{i(2 \phi _{B_{\bot}} +\phi_{\Pi})}}{\left| 1-2Re^{i(2 \phi _{B_{\bot}} +\phi_{\Pi})}\right|^2}\right)\frac{\mathcal{B}_{\bot}}{D_{gs}}, e^{i 2 \phi _{B_{\bot}}}\frac{1-2Re^{-i(2 \phi _{B_{\bot}} +\phi_{\Pi})}}{\left| 1-2Re^{-i(2 \phi _{B_{\bot}} +\phi_{\Pi})}\right|^2} \right)
 \end{split}
\end{equation}
\end{widetext}
It is interesting to discuss the limit case $R \ll1$, i.e., when $\cal{B_{\bot}}$ is predominant. In this case, the eigenstates are, after collecting a global phase factor $e^{i  \phi _{B_{\bot}}}$
\begin{equation}
\label{eq:eigenvectors_approx_B}
\begin{split} 
\ket{0} &= \left(e^{i \phi _{B_{\bot}}}\frac{\mathcal{B}_{\bot}}{D_{gs}},-\sqrt{2}, e^{i \phi _{B_{\bot}}}\frac{\mathcal{B}_{\bot}}{D_{gs}}\right)=\\
&=-\sqrt{2}\left( \ket{S_z=0} - \frac{\mathcal{B}_{\bot}}{2D_{gs}}\ket{+}_{\phi _{B_{\bot}}} \right)\\
  \ket{1} &= \left( e^{-i  \phi _{B_{\bot}}}, 0, -e^{i  \phi _{B_{\bot}}}\right)= \sqrt{2} \ket{-}_{\phi _{B_{\bot}}}\\
 \ket{2} &= \left( e^{-i  \phi _{B_{\bot}}}, \sqrt{2}\frac{\mathcal{B}_{\bot}}{D_{gs}}, +e^{i  \phi _{B_{\bot}}}\right)= \\
 &=\sqrt{2}\left( \ket{+}_{\phi _{B_{\bot}}}+ \frac{\mathcal{B}_{\bot}}{2D_{gs}}\ket{S_z=0}\right)\ \\
 \end{split}
\end{equation}
where we introduced the vectors: \\ 
\begin{equation}
\label{eq:eigenvectors_approx_B_rotation}
\begin{split} 
 \ket{-}_{\phi _{B_{\bot}}} &= \frac{1}{\sqrt2}\left( e^{-i  \phi _{B_{\bot}}}, 0, -e^{i  \phi _{B_{\bot}}}\right)=\\
 & =\frac{1}{\sqrt2}\left(e^{-i  \phi _{B_{\bot}}}\ket{S_z=+1} -e^{i  \phi _{B_{\bot}}}\ket{S_z=-1} \right)\\
 \ket{+}_{\phi _{B_{\bot}}}&= \frac{1}{\sqrt2}\left( e^{-i  \phi _{B_{\bot}}}, 0, +e^{i  \phi _{B_{\bot}}}\right)=\\
&=\frac{1}{\sqrt2}\left(e^{-i  \phi _{B_{\bot}}}\ket{S_z=+1} +e^{i  \phi _{B_{\bot}}}\ket{S_z=-1} \right)\\
\end{split}
\end{equation}
Considering  Eq.s (\ref{eq:eigenvectors_approx_B}),  it is clear that for $R<<1$ the effect of the orthogonal magnetic field is: \\
\begin{itemize}
    \item to rotate the original dressed states $\ket{-}$ and $\ket{+}$ of angle $\phi _{B_{\bot}}$ equal to field orientation
    \item create a small mixing between $\ket{+}_{\phi _{B_{\bot}}}$ and $\ket{S_z=0}$ state.
\end{itemize}
These results are usually derived using second-order degenerate perturbation theory.
Now, let's turn our attention to the general case $R\neq0$. If we introduce the angle
\begin{equation}
\begin{split}
\label{eq:theta_bis}
\theta &=\frac{1}{2}\arg\left(e^{i2 \phi _{B_{\bot}}}\frac{1-2Re^{-i(2 \phi _{B_{\bot}} +\phi_{\Pi})}}{\left| 1-2Re^{-i(2 \phi _{B_{\bot}} +\phi_{\Pi})}\right|^2}\right)= \\
&=\frac{1}{2}\arg\left(e^{2i \phi _{B_{\bot}}}-2Re^{-i \phi _{\Pi}}\right)
\end{split}
\end{equation}
and we collect a phase factor $e^{i  \theta}$ in Eq.s (\ref{eq:eigenvectors_approx}) we obtain:
\clearpage
\begin{widetext}
\begin{equation}
\label{eq:eigenvectors}
\begin{split} 
& \ket{0} =-\sqrt{2}\left[e^{i  (\phi _{B_{\bot}}-\theta)} \ket{S_z=0} - \frac{\mathcal{B}_{\bot}}{2D_{gs}}\left( \cos(\phi _{B_{\bot}}-\theta)\ket{+}_{{\theta}}+ i\sin(\phi _{B_{\bot}}-\theta)\ket{-}_{{\theta}}  \right) \right] \\
& \ket{1} = \sqrt{2}\left( \ket{-}_{\theta}+ \frac{\mathcal{B}_{\bot}}{2D_{gs}}i\sin(\phi _{B_{\bot}}-\theta)\ket{S_z=0}\right)\\
& \ket{2} = \sqrt{2}\left( \ket{+}_{\theta}+ \frac{\mathcal{B}_{\bot}}{2D_{gs}}\cos(\phi _{B_{\bot}}-\theta)\ket{S_z=0}\right)\\
\end{split}
\end{equation}
\end{widetext}
Eq. (\ref{eq:eigenvectors}) tell us that the effect of orthogonal magnetic and electric field is to rotate the original dressed states $\ket{-}$ and $\ket{+}$ of an angle $\theta$. Thus, we can rename the eigenergies as:
\begin{equation}
\label{eq:eigenvalues_approx_renamed}
\begin{split}
&E_{-,\theta}=E_1 \\
&E_{+,\theta}=E_2
\end{split}
\end{equation}
\section{Effect of an axial magnetic field}
We have shown in the previous sections that the combined contribution of a total electric field and a weak orthogonal field can be summarized in three effects:
\begin{itemize}
	\item Rotation of the original dressed states  $\ket{S_z=0},\ket{-},\ket{+}$ of an angle $\theta$ along the $z$-axis, giving the eigenstates $\ket{S_z=0}, \ket{-}_{\theta}, \ket{+}_{\theta}$. This is valid if we neglegt the small mixing between $\ket{-}_{\theta},\ket{+}_{\theta}$ and $\ket{S_z=0}_{\theta}$
	\item Creation of an energy gap between $\ket{+}_{\theta}$ and $\ket{-}_{\theta}$, $E_{gap}=2\left[ \left(\frac{\mathcal{B}_{\bot}^2 }{2D_{gs}}\right)^2 + \mathcal{E}^2 -  \mathcal{E}\frac{\mathcal{B}_{\bot}^2 }{D_{gs}}\cos(2\phi_{B_{\bot}}) \right]^\frac{1}{2}$
	\item Decrease in energy of the $\ket{0}$ state to $E_0= - \frac{\mathcal{B}_{\bot}^2}{D_{gs}}< 0$
\end{itemize}

Considering negligible the mixing between $\ket{-}_{\theta},\ket{+}_{\theta}$ and $\ket{S_z=0}$, we can rewrite the Hamiltonian of the system as:

\begin{equation}
\label{eq:Ham_matrix_B_z}
H_{B_{\parallel}=0}=
\begin{pmatrix}
E_0-E_m            & 0                             & 0                         \\
0                         & - \frac{E_{gap}}{2}  & 0                           \\
0                         & 0                             &  + \frac{E_{gap}}{2}\\ 
\end{pmatrix}
\end{equation}

in the basis $\ket{0}_{\theta},\ket{-}_{\theta}, \ket{+}_{\theta}$, where $E_m=\frac{E_{+,\theta} + E_{-,\theta}}{2}=D_{gs} + \frac{\mathcal{B}_{\bot}^2}{2D_{gs}}$ was chosen as the zero of the energy scale.

Following Ref. \cite{jamonneau2016competition} if we consider a magnetic field $B_{\parallel}$  aligned along the NV axis, the total Hamiltonian will be

\begin{equation}
\label{eq:Ham_matrix_B_z_bis}
H_{B_{\parallel}\neq0}=
\begin{pmatrix}
E_0-E_m            &                                 & 0                               \\
0                         & - \frac{E_{gap}}{2}   & \mathcal{B}_{\parallel}           \\
0                         & \mathcal{B}_{\parallel}             &  + \frac{E_{gap}}{2}   \\ 
\end{pmatrix}
\end{equation}

where $\mathcal{B}_{\parallel} = g\mu_BB_{\parallel}$. If we consider only the subspace $\ket{-}_{\theta},\ket{+}_{\theta}$ the Hamiltonian becomes

\begin{equation}
\label{eq:Ham_matrix_pm}
H_{red}=
\begin{pmatrix}
-\frac{E_{gap}}{2} & \mathcal{B}_{\parallel}\\
\mathcal{B}_{\parallel}          & +\frac{E_{gap}}{2}\\
\end{pmatrix}
\end{equation}

If we now change the basis to  $\ket{S_z=-1}_{\theta},\ket{S_z=+1}_{\theta}$ we have:


\begin{equation}
\label{eq:Ham_matrix_reduced_pm_1}
H_{red}=
\begin{pmatrix}
-\mathcal{B}_{\parallel} & \frac{E_{gap}}{2} \\
 \frac{E_{gap}}{2} & \mathcal{B}_{\parallel}.\\
\end{pmatrix}
\end{equation}

This Hamiltonian can be diagonalized exactly, considering the system as one with effective spin $s=\frac{1}{2}$. The resulting eigenvectors are:
\begin{equation}
\label{eq:eigenvectors_comb_B_z_bis}
\begin{split} 
&\ket{-}_{\theta, \mathcal{B}_{\parallel}} = \sin(\frac{\gamma}{2})e^{-i\theta}\ket{S_z=+1} - \cos(\frac{\gamma}{2})e^{i\theta}\ket{S_z=-1} \\
& \ket{+}_{\theta, \mathcal{B}_{\parallel}} = \cos(\frac{\gamma}{2})e^{-i\theta}\ket{S_z=+1} + \sin(\frac{\gamma}{2})e^{i\theta}\ket{S_z=-1}
 \end{split}
\end{equation}

with 
\begin{equation}
\label{eq:gamma}
\tan{\gamma} = \frac{E_{gap}/2}{\mathcal{B}_{\parallel}}.
\end{equation}

The corresponding eigenenergies are:

\begin{widetext}
\begin{equation}
\label{eq:eigenvalues_comb_B_z_bis}
\begin{split}
& E_{0'}^{(2)} = - \frac{\mathcal{B}_{\bot}^2}{2D_{gs}} \\
& E_{-,\theta, \mathcal{B}_{\parallel}}^{(2)} = E_m - \sqrt{(E_{gap}/2)^2 + \mathcal{B}_{\parallel}^2} = D_{gs} + \frac{\mathcal{B}_{\bot}^2}{2D_{gs}} - \sqrt{(E_{gap}/2)^2 + \mathcal{B}_{\parallel}^2}\\
& E_{+,\theta, \mathcal{B}_{\parallel}}^{(2)} = E_m + \sqrt{(E_{gap}/2)^2 + \mathcal{B}_{\parallel}^2} = D_{gs} + \frac{\mathcal{B}_{\bot}^2}{2D_{gs}} + \sqrt{(E_{gap}/2)^2 + \mathcal{B}_{\parallel}^2} \\
\end{split}
\end{equation}
\end{widetext}

We underline that the states $\ket{-}_{\theta, \mathcal{B}_{\parallel}}$, $\ket{+}_{\theta, \mathcal{B}_{\parallel}}$  are in general a  \underline{non- } balanced superposition of high field states $\ket{S_z=+1},\ket{S_z=+1}$.  The dependence of the coefficients of high field states as function of the axial field is shown in Figure \ref{fig:coefficient_high_field} for the $\ket{+}_{\theta, \mathcal{B}_{\parallel}}$ where we consider $0<\gamma<\pi$. For $\mathcal{B}_{\parallel}=0$, $\gamma=\frac{\pi}{2}$ the coefficients are equal in modulus. Increasing $\mathcal{B}_{\parallel}$ to higher positive values, we have that $\gamma<\frac{\pi}{2}$ and the coefficient relative to the state $\ket{S_z=+1}$ becomes larger than the one relative to $\ket{S_z=1}$. Decreasing $\mathcal{B}_{\parallel}$ to lower positive values, the coefficient relative to the state $\ket{S_z=-1}$ becomes larger than the one relative to $\ket{S_z=+1}$. This explains the transition from dressed states to partially dressed states and then to quasi-high-field states.

The expectation values of the axial spin are:
\begin{equation}
\label{eq:expectation_values_B_z}
\begin{split}
&  _{\theta, \mathcal{B}_{\parallel}}\langle +| S_z | +   \rangle_{\theta, \mathcal{B}_{\parallel}} = \cos^2\left(\frac{\gamma}{2}\right) - \sin^2\left(\frac{\gamma}{2}\right) = \cos(\gamma)\\
& _{\theta, \mathcal{B}_{\parallel}}\langle -| S_z | -  \rangle_{\theta, \mathcal{B}_z} = \sin^2\left(\frac{\gamma}{2}\right) -\cos^2\left(\frac{\gamma}{2}\right)= -\cos(\gamma)\\
& \langle 0| S_z | 0   \rangle_{\mathcal{B}_{\parallel}} = 0.
\end{split}
\end{equation}

From this expression it follows that partially dressed states for opposite values of axial field $B_{\parallel}$ have opposite values of $\langle S_z \rangle$ because $\cos(\frac{\pi}{2}-\gamma)=-\cos(\frac{\pi}{2}-\gamma)$, where, again, we recall that $\gamma=\frac{\pi}{2}$ for $\mathcal{B}_{\parallel}=0$.

\begin{figure*}[ht]
\includegraphics[width=\textwidth]{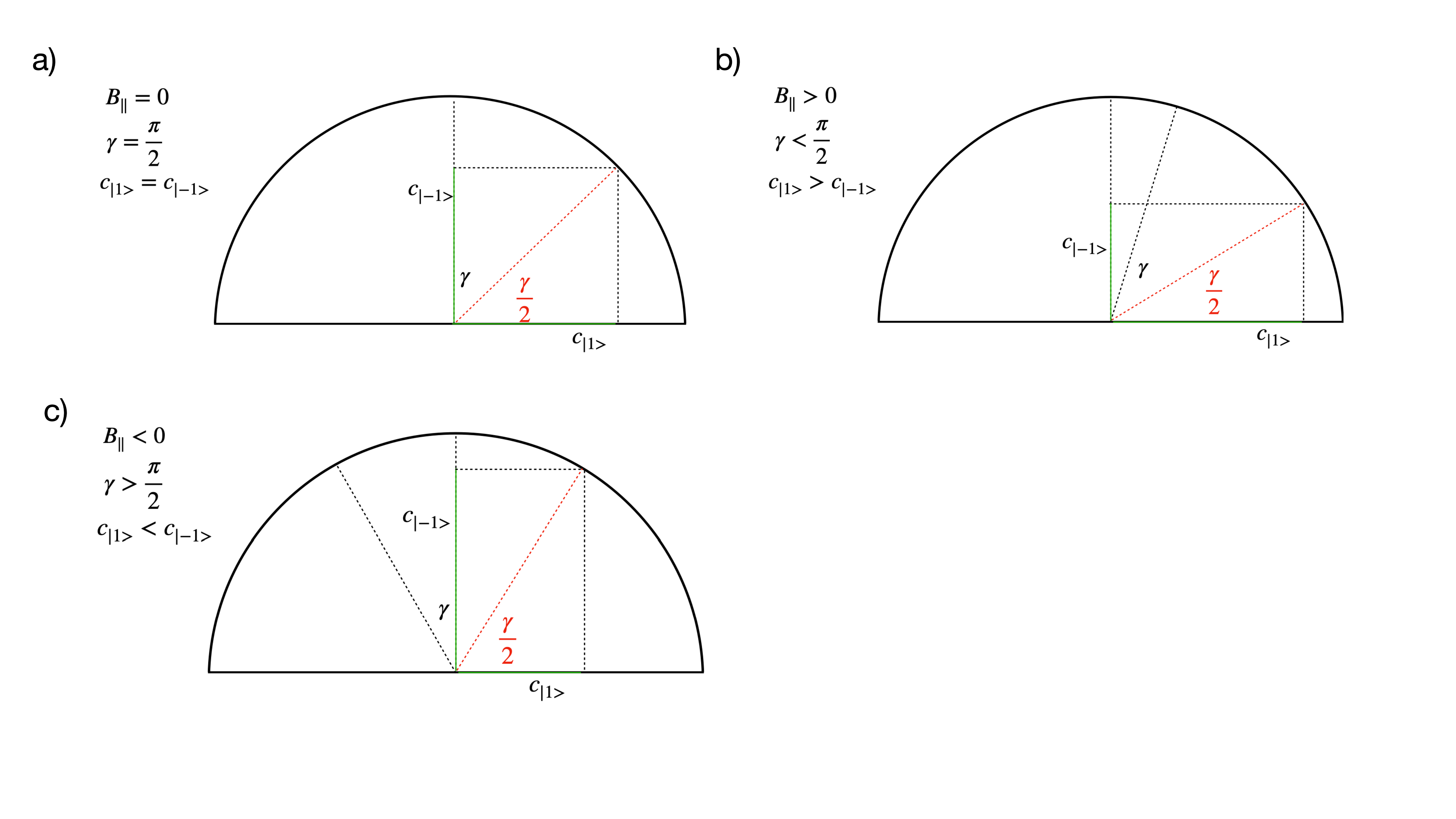}
\caption{Different balance of coefficients of high-field for a) $B_{\parallel}=0$ b) $B_{\parallel}>0$ c) $B_{\parallel}<0$. For $B_{\parallel}=0$ the coefficients are equal, and the superposition is balanced, corresponding to dressed states. For $B_{\parallel}>0$ the superposition is unbalanced toward $\ket{S_z=+1}$ and for$B_{\parallel}<0$ the superposition is unbalanced toward $\ket{S_z=-1}$ }
\label{fig:coefficient_high_field} 
\end{figure*}

Let's now consider the mixing between  $\ket{-}_{\theta},\ket{+}_{\theta}$ and $\ket{S_z=0}$, that has not been considered so far, since we intend to prove that it is negligible. The basis to be considered is the one formed by the eigenvectors $\ket{0}, \ket{1}, \ket{2}$ of Eq.s (\ref{eq:eigenvectors}). On this basis the total Hamiltonian is: \\

\begin{widetext}    
\begin{equation}
\label{eq:Ham_matrix_B_z_third_order}
H_{B_z\neq0}=
\begin{pmatrix}
E_0-E_m                                                                                                     &  - \frac{\mathcal{B}_{\parallel} \mathcal{B}_{\bot}}{2D_{gs}} \cos(\phi _{B_{\bot}}-\theta)                               &  +i \frac{\mathcal{B}_{\parallel} \mathcal{B}_{\bot}}{2D_{gs}}\sin(\phi _{B_{\bot}}-\theta)                            \\
- \frac{\mathcal{B}_{\parallel} \mathcal{B}_{\bot}}{2D_{gs}}\cos(\phi _{B_{\bot}}-\theta)                          & - \frac{E_{gap}}{2}                                                                                              & \mathcal{B}_{\parallel}           \\
-i \frac{\mathcal{B}_{\parallel} \mathcal{B}_{\bot}}{2D_{gs}}\sin(\phi _{B_{\bot}}-\theta)                          & \mathcal{B}_{\parallel}                                                                                                      &  + \frac{E_{gap}}{2}   \\ 
\end{pmatrix}
\end{equation}
\end{widetext}


Considering our experimental conditions, $\frac{\mathcal{B}_{\parallel} \mathcal{B}_{\bot}}{2D_{gs}}i\sin(\phi _{B_{\bot}}-\theta) \ll E_0-E_m = D_{gs} + \frac{\mathcal{B}_{\bot}^2}{D_{gs}}$ and we can consider the term $\frac{\mathcal{B}_{\parallel} \mathcal{B}_{\bot}}{2D_{gs}}$ as a perturbation of the Hamiltonian in absence of mixing. Without going into detailed calculations, from perturbation theory, we know that the effect of the perturbation is of the order of $\frac{\mathcal{B}_{\parallel}^2 \mathcal{B}_{\bot}^2}{D_{gs}^3}$ on the eigenergies and can be neglected, that was our initial assumption.

\section{Polarization}
\label{Calculation Polarization}

This section will study how to excite dressed and partially dressed states using an oscillating magnetic field  $\vec{B_{osc}}$ with suitable polarization. \\

\subsection{Dressed states}

We consider the effect of an oscillating magnetic field $\vec{B}_{osc}$ in driving the transition between the state $\ket{0}$ and dressed states ($\ket{+}_{\theta} \,,\, \ket{-}_{\theta}$). If the magnetic field is aligned along the direction defined by  $\theta$ and the corresponding versor $\hat{v}$, its expression is:

\begin{equation}
\begin{split}
\vec{B}_{osc} & = B_{\theta} \cos(\omega_d t) \hat{v}\\
 & =B_{\theta} \cos(\omega_d t) \left(\cos\theta \hat{x} + \sin\theta\hat{y}\right)
 \end{split}
\end{equation}

where $B_{\theta}$ is the amplitude of the field along $\theta$, and $\omega_d$ is the angular frequency.
The time-dependent Hamiltonian is

\begin{equation}
\label{eq:Ham_polarization}
H_{0,dressed} = D_{gs}S_{z}^{2} + 2\Omega_{\theta}\cos(\omega_d t)\left( \cos\theta S_{x}  + \sin\theta S_{y} \right)
\end{equation}

where $2\Omega_{\theta}= g \mu_B B_{\theta} $
For simplicity, we have considered only the part due to zero-field-splitting $D$. In the interaction description defined by $H_{0}^{(1)}=\omega_dS_z^2$, we have

\begin{widetext}
\begingroup
\setlength\arraycolsep{1pt}
\begin{equation}
\begin{split}
H_{1;dressed} & = U_0^{(1)}(t)^{\dagger}H_{0,dressed}U_0^{(1)}(t) -i U_0^{(1)}(t)^{\dagger} \partial_t U_0^{(1)}(t) \\
=& 
\begin{bmatrix}
D_{gs}-\omega_d                                                                                               &  \sqrt{2}\Omega e^{i\omega_dt} \cos(\omega_d t) (\cos\theta -i \sin \theta)                              &   0                            \\
\sqrt{2}\Omega e^{-i\omega_dt} \cos(\omega_d t) (\cos\theta + i \sin \theta)                               & 0                                                                                                                &\sqrt{2}\Omega e^{-i\omega_dt} (\cos\theta -i \sin \theta) \cos(\omega_d t)           \\
0                                                                                                                 &  \sqrt{2}\Omega e^{i\omega_dt} \cos(\omega_d t) (\cos\theta + i \sin \theta)                             & D_{gs} -\omega_d \\ 
\end{bmatrix}\\
\approx&\begin{bmatrix}
D_{gs}-\omega_d                                                                     &  \frac{1}{\sqrt{2}}\Omega e^{-i \theta}                                                                                                      &    0                                             \\
 \frac{1}{\sqrt{2}}\Omega e^{+i \theta}                                                        & 0                                                                                                                                    & \frac{1}{\sqrt{2}}\Omega  e^{-i \theta}       \\
0                                                                                       & \frac{1}{\sqrt{2}}\Omega e^{+i \theta}                                                                                                      & D_{gs} -\omega_d \\ 
\end{bmatrix}\\
\end{split}
\end{equation}
\endgroup
\end{widetext}

Where $U_0^{(1)}(t) = \exp \left(-i\omega_dS_z^2 \right)$, and we applied the rotating wave approximation, neglecting the fast terms rotating at $2\omega_dt$ in the second row. From the last row, it is clear that when  $\vec{B}_{osc}$ is aligned with $\theta$, it induces transition between $\ket{0}$ and $\ket{+}_{\theta}$. Analogously, when $\vec{B}_{osc}$ is aligned orthogonal to $\theta$ it induce transitions between  $\ket{0}$ and $\ket{-}_{\theta}$.

\subsection{Partially Dressed states}
Let's assume now an elliptically polarized MW excitation with the major axis aligned along $\theta$:

\begin{widetext}
\begin{equation}
\begin{split}
\vec{B}_{osc} & =B_{\theta} \cos(\omega_d t) \hat{v}  + B_{\theta +\pi/2} \sin(\omega_d t) \hat{w}\\
 & =B_{\theta} \cos(\omega_d t) \left[\cos\theta \hat{x} + \sin\theta\hat{y}\right] + B_{\theta +\pi/2} \cos(\omega_d t) \left[-\sin\theta\hat{x} + \cos\theta\hat{y}\right] \\
\end{split}
\end{equation}
\end{widetext}
where $B_{\theta}$ is the amplitude of the field along $\theta$, $\hat{v}$ is the versor defined by $\theta$, $B_{\theta + \pi/2}$ is the amplitude of the field along $\theta + \pi/2$, $\hat{w}$ is the versor defined by $\theta + \pi/2$, with $B_{\theta} > B_{\theta+\pi/2}$. If $B_{\theta} = B_{\theta+\pi/2}$, the magnetic field is circularly clockwise polarized.
The total time-dependent Hamiltonian is
\begin{equation}
	\begin{split}
	\label{eq:Ham_polarization_part}
	 &H_{0,p-dressed} = \\ &D_{gs}S_{z}^{2} + 2\Omega_{\theta}\cos(\omega_d t)\left( S_{x}\cos\theta  + S_{y}\sin\theta \right) +\\ &2\Omega_{\theta+\pi/2}\sin(\omega_d t)\left(S_{x}(-\sin\theta) + S_{y}\cos \theta \right)
	 \end{split}
\end{equation} \\
where $2\Omega_{\theta}= g \mu_B B_{\theta} $ and $2\Omega_{\theta+\pi/2}= g \mu_B B_{\theta+\pi/2} $.\\ Also in this case, we have considered only the part due to zero-field-splitting $D_{gs}$. We move to the interaction description defined by $H_{0}^{(1)}=\omega_dS_z^2$
\begin{widetext}
\begingroup
\setlength\arraycolsep{1pt}
\begin{equation}
\begin{split}
H_{1,p-dressed} & = U_0^{(1)}(t)^{\dagger}H_{0,p-dressed}U_0^{(1)}(t) -i U_0^{(1)}(t)^{\dagger} \partial_t U_0^{(1)}(t) \\
=&\begin{bmatrix}
D_{gs}-\omega_d                                                                                                                                           &  \frac{1}{\sqrt{2}}\left(\Omega_{\theta} +\Omega_{\theta + \pi/2 }\right) e^{-i \theta}                                                                                                    &    0                                             \\
\frac{1}{\sqrt{2}}\left(\Omega_{\theta} +\Omega_{\theta + \pi/2 }\right) e^{+i \theta}                                                       & 0                                                                                                                                    & \frac{1}{\sqrt{2}}\left(\Omega_{\theta} -\Omega_{\theta + \pi/2 }\right) e^{-i \theta}       \\
0                                                                                       & \frac{1}{\sqrt{2}}\left(\Omega_{\theta} -\Omega_{\theta + \pi/2 }\right) e^{+i \theta}                                                                                                    & D_{gs} -\omega_d \\ 
\end{bmatrix}\\
\label{eq:Ham_polarization_part_inter}
\end{split}
\end{equation}
\endgroup
\end{widetext}

From the  last row of Eq. (\ref{eq:Ham_polarization_part_inter}), it can be seen that the elliptically polarized MW excitation drive transition between the states $\ket{0}$ and $\ket{+}_{\theta, \mathcal{B}_{\parallel}}$ in Eq. (\ref{eq:eigenvectors_comb_B_z_bis}), if:

\begin{equation}
\begin{split}
\cos(\frac{\gamma}{2}) = \frac{1}{\sqrt{2}}\frac{\Omega_{\theta} +\Omega_{\theta + \pi/2 }}{\sqrt{\Omega_{\theta}^2 +\Omega_{\theta + \pi/2 }^2}}\\
\sin(\frac{\gamma}{2})= \frac{1}{\sqrt{2}}\frac{\Omega_{\theta} -\Omega_{\theta + \pi/2 }}{\sqrt{\Omega_{\theta}^2 +\Omega_{\theta + \pi/2 }^2}}\\
\end{split}
\end{equation}


Increasing $B_{\parallel}$, the angle $\gamma$ goes to $0$ (see Eq. (\ref{eq:gamma})), we have $\Omega_{\theta}=\Omega_{\theta + \pi/2 }$ and we recover the well know results that transition between $\ket{0}$ and strong axial-field states are driven by circularly polarized oscillating magnetic field \cite{pellicer2024versatile}. Similarly, it can be shown that when $B_{osc}$ is aligned with the mayor-axis orthogonal to $\theta$ it induces transition between  $\ket{0}$ and $\ket{-}_{\theta,\mathcal{B}_{\parallel}}$. 

Summarizing, i) transitions between $\ket{0}$ and dressed states are promoted by the linearly polarized oscillating magnetic field, aligned along $\theta$, i.e. along the direction determined by the competition between orthogonal magnetic field and total orthogonal electric field, or orthogonal to this direction ii) transitions between $\ket{0}$ and partially dressed states are promoted by elliptically polarized oscillating magnetic field aligned or orthogonal to $\theta$ iii) Furthermore, it is well known that transitions between $\ket{0}$ and strong-axial field states are promoted by circularly polarized oscillating magnetic field \cite{pellicer2024versatile}.

\section{Decoherence in a Free Induction Decay Measurement for a single NV center}
\label{Calculation dec single}

We use a derivation similar to the one presented in \cite{bauch2020decoherence,bauch2018ultralong, dobrovitski2008decoherence,jamonneau2016competition}. We first consider the case of a single NV, and then discuss how to generalize the discussion to an ensemble of NV's.\\
The sources of decoherence for a single NV center are:
\begin{enumerate}
	\item Coupling with $^{13}$C 
	\item Coupling with $^{14}$N or $^{15}$N
	\item Coupling with other spins
	\item Temporal fluctuations of external fields
\end{enumerate}
Decoherence due to the bath of surrounding spins, points 1-3 in the previous list, is a pure quantum phenomenon due to the entanglement of the NV center with the surrounding spin bath of |\cite{zurek2003decoherence, dobrovitski2008decoherence}. For a limited class of problems, with so-called "nonbranching" evolution, we can map the original quantum spin bath onto a classical random magnetic field $\vec{B}_{s-dec}$ \cite{bauch2020decoherence}. We can consider this field as aligned along the z-axis, $\vec{B}_{s-dec}=B_{s-rand}\vec{z}$, since the orthogonal part of this field is due to flip-flop terms in the spin-spin coupling that are suppressed by the big zero-field splitting $D_{gs}$. 

The temporal fluctuations of the external fields (point 4 in the previous list), can be described by two other stochastic magnetic $\vec{B}_{t-dec}$ and total electric fields $\vec{\Pi}_{t-dec}$. Finally, we consider that the effect of the spin-spin coupling and of the temporal fluctuations adds up, giving a total stochastic magnetic field $\vec{B}_{dec}=\vec{B}_{s-dec} + \vec{B}_{t-dec}$. 

For simplicity, we consider $\vec{B}_{dec}$ and $\vec{\Pi}_{dec}$ have a Gaussian distribution with zero mean and standard deviation $\sigma_{B_{dec,i}}$ and $\sigma_{\Pi_{dec,i}}$ in each cartesian direction $i$.

The fluctuations of the stochastic fields $\vec{B}_{dec}$ and  $\vec{\Pi}_{dec}$ represented by the random variable $\delta \vec{B}_{dec}$ and  $\delta\vec{\Pi}_{dec}$ induce fluctuations in the eigenenergies of the NV center and consequently fluctuations $\delta \nu$ in the ODMR transition frequencies. Fluctuations $\delta \nu$ in the ODMR transition frequencies will induce a decay in FID signal \cite{jamonneau2016competition}. To fix the idea let's focus on the $\ket{0}\rightarrow\ket{-}_{\theta, \mathcal{B}_{\parallel}}$ ODMR transition. In general, the FID signal is proportional to the probability $p_{\ket{0}}$ to be in the $\ket{0}$

\begin{equation} 
p_{\ket{0}}(\tau)=\frac{\left[1- \cos(\phi + \delta \phi )\right]}{2}
\label{eq:PL_app}
\end{equation}

where $\tau$ is the duration of the free precession interval, $\delta\phi=\int_0^{\tau}2\pi\delta\nu dt$ and $\phi= 2\pi\Delta \tau$ are respectively the stochastic and static phase acquired during the free precession interval, and $\Delta= \nu_{MW}-\nu_{-,\theta, \mathcal{B}_{\parallel}}$ is the detuning between the microwave excitation frequency $\nu_{MW}$ and the ODMR transition. If we now consider the average on the different experimental realizations 

\begin{equation} 
 p_{FID}(\tau)=\langle p_{\ket{0}}(\tau) \rangle =\frac{\left[1-e^{\langle \delta \phi ^2\rangle/2}\cos 2\pi \Delta \tau\right] }{2}
 \label{eq:PLbis}
\end{equation}

we see that the FID decay is determined by the variance of the stochastic phase acquired during the free precession interval, $\langle \delta \phi ^2\rangle$, where;

\begin{equation} 
\label{eq:phinu}
 \langle \delta \phi ^2\rangle= 4\pi^2\int_0^{\tau}dt\int_0^{\tau}dt'\langle \delta \nu(t)\delta \nu(t')\rangle.
\end{equation}

and $\langle \delta \nu(t)\delta \nu(t')\rangle$ is the correlation function. To evaluate $\langle \delta \nu(t)\delta \nu(t')\rangle$ we consider:
\begin{enumerate}
        \item $\vec{\delta B_{dec}}$ and  $\vec{\delta\Pi_{dec}}$ random variables with correlations functions dacaying exponentially.  
\begin{equation}
\label{eq:corrfunc}
\begin{split}
&\langle \delta B_{dec,i}(t)\delta B_{dec,i}(t')_i \rangle=\sigma_{B_{dec.i}}^2  e^{|t-t'|/\tau_{c,B_{dec,i}}} \\
&\langle \delta \Pi_{dec,i}(t)\delta \Pi_{dec,i}(t')_i \rangle=\sigma_{\Pi_{dec,i}}^2  e^{|t-t'|/\tau_{c,\Pi_{dec,i}}} 
 \end{split}
  \end{equation}
 where $\delta B_{dec,i}$ and $\delta \Pi_{dec,i}$ are the component of fluctuations of $\vec(B)_dec$ and $\vec(\Pi)_dec$ along the i-direction, with $i=x,y,z$
	\item $\delta \nu_{res}$ dependence  on $\delta B_{dec,i}$ and  $\delta\Pi_{dec,i}$  is set by the value of the static fields $\vec{B}$ and $\vec{\Pi}$. Considering $\sigma_{B_{dec,i}}= \sqrt{{\langle \delta B_{dec,i} ^2 \rangle}} \ll B_i$ and $\sigma_{\Pi_{dec,i}}= \sqrt{{\langle \delta  \Pi_{dec,i} ^2} \rangle} \ll \Pi_i , \forall i$, $\delta B_{dec}$ and  $\delta\Pi_{dec}$ act as perturbations on the eigenstates and eigenvalues set by $\vec{B}$ and $\vec{\Pi}$ in the Hamiltonian in Eq. (\ref{eq:Ham_matrix_B_z_bis}).  
\end{enumerate}

Now we will study the different scenarios set by the values of the static fields $\vec{B}$ and $\vec{\Pi}$

\subsection{Limit case: Weak axial field}

To simplify the discussion, let us consider the quantities introduced before: 

\begin{equation}
\mathcal{B}_{\bot} =\frac{g\mu_B}{h}B_{\perp} \ \ \text{,} \ \ \mathcal{E}=d_{\bot}\Pi_{\bot} \ \ \text{,} \ \ \mathcal{B}_{\parallel} =\frac{g\mu_B}{h}B_{\parallel}.
\end{equation}

When the condition 

\begin{equation}
\sqrt{\mathcal{E}^2 + \frac{\mathcal{B}_{\bot}^2}{D_{gs}}} \gg \mathcal{B}_{\parallel}
\end{equation}

is satisfied, the eigenstates are dressed states $\ket{0}, \ket{+}_{\theta}, \ket{-}_{\theta}$, see Eq.s \ref{eq:eigenvectors_rotated}. The eigenenergies are $E_0, E_{+,\theta}, E_{-,\theta}$ see Eq.s \ref{eq:eigenvalues_approx} and \ref{eq:eigenvalues_approx_renamed}, with 

\begin{equation}
\label{eq:energy_gap}
E_{gap}= E_{+,\theta} - E_{-,\theta} = 2\left[ \left(\frac{\mathcal{B}_{\bot}^2 }{2D_{gs}}\right)^2 + \mathcal{E}^2 -  \mathcal{E}\frac{\mathcal{B}_{\bot}^2 }{D_{gs}}\cos(2\phi) \right]^\frac{1}{2}
\end{equation}

 For a given $\theta$, the angle $\phi = -2\theta + \pi/2$ defines a preferred direction $x'$ for a fluctuation of the total electric field $\delta\Pi_{dec,x'}$. Along this direction, the contribution of the fluctuations of the total electric field, $\delta\Pi_{dec,x'}$ enters at first order in the perturbation. Hence, the fluctuations $\delta B_{dec,i}$ and $\delta \Pi_{dec,i}$ induce the following fluctuations in the eigenenergies, considering till the second order:

\begin{equation}
\label{eq:fluctuations_dressed}
\begin{split}
& \delta E_{+,\theta} = - d_{\bot}\delta\Pi_{dec,x'} + \frac{(g\mu_B\delta B_{dec,z})^2}{E_{gap}} + \frac{(d_{\bot}\delta\Pi_{dec,y'})^2}{E_{gap}} +\\
& + \frac{(g\mu_B\delta B_{dec,x'})^2}{D_{gs}} \\
& \delta E_{-,\theta} = +d_{\bot}\delta\Pi_{dec,x'} + \frac{(g\mu_B\delta B_{dec,z})^2}{E_{gap}} + \frac{(d_{\bot}\delta\Pi_{dec,x'})^2}{E_{gap}} + \\
& + \frac{(g\mu_B\delta B_{dec,y'})^2}{D_{gs}} \\
& \delta E_{0} = -\frac{(g\mu_B\delta B_{dec,x'})^2}{D_{gs}} -\frac{(g\mu_B\delta B_{dec,y'})^2}{D_{gs}}
\end{split}
\end{equation}

Where $y'$ is the direction orthogonal to $x'$. Considering the $\ket{0}\rightarrow\ket{-}_{\theta}$ ODMR transition, we have the following expression for the fluctuation of the resonance frequency: 

\begin{equation}
\label{eq:fl}
 \delta\nu_- = \delta E_{-,\theta}  - \delta E_{0}  = d_{\bot}\delta\Pi_{dec,x'} + \frac{(g\mu_B\delta B_{dec,z})^2}{E_{gap}} 
\end{equation}

where we neglected the term proportional to $\delta B_{dec,x'} $ because $D_{gs}\gg g \mu_B\delta B_{dec,x'}$, and we neglected the term proportional to $\delta\Pi_{dec,y'}$ because we consider $\vec{\Pi_{dec}}$ isotropic.

We underline that the effect of fluctuations of the axial magnetic field $\delta B_z$ is only at the second order, and it is scaled by the value of the energy gap $E_{gap}$. This is because $\ket{-}_{\theta}$ is a dressed state with $\langle S_z \rangle=0$. 

If we now consider a slowly fluctuating bath with $\tau \ll \tau_{c,B_{dec,z}}, \tau_{c,\Pi_{dec,y}} $,  using the expression in Eq. ($\ref{eq:fl}$) through Eq. ($\ref{eq:corrfunc}$) and ($\ref{eq:phinu}$), we can calculate the resulting variance of the phase fluctuation as

\begin{equation} 
\label{eq:phi_tau}
\langle \delta \phi ^2\rangle = 4\pi^2\left[d_{\bot}^2\sigma_{\Pi_{dec,x'}}^2+\frac{(g\mu_B)^4\sigma_{B_{dec,z}}^4}{E_{gap}^2}\right]\tau^2
\end{equation}

and so, from Eq. (\ref{eq:PLbis}), the FID decay: 

\begin{equation} 
\label{eq:phi}
 p_{FID}(\tau) = \langle p_{\ket{0}(\tau)} \rangle = \frac{\left[1-e^{-\left(\frac{\tau}{T_2^*}\right)^2}\cos 2\pi \Delta \tau\right] }{2},
\end{equation}

with:

\begin{equation} 
\label{eq:coherence_time_dressed}
T_{2,dressed}^* = \frac{1}{\sqrt{2}\pi}\frac{1}{\sqrt{d_{\bot}^2\sigma_{\Pi_{dec,x'}}^2+\frac{(g\mu_B)^4\sigma_{B_{dec,z}}^4}{E_{gap}^2}}},
\end{equation}

$T_2^*$ defines the characteristic time-scale of the FID decay. In this scenario, the more effective decoherence source is $\Pi_{dec,x'}$, i.e., the temporal fluctuations of the electrical fields or of the strain along $y'$ (the direction defined by $\phi = -2\theta + \pi/2$).

\subsection{Limit case: strong axial field}

If we now consider the opposite condition, i.e.: 

\begin{equation}
\mathcal{B}_z \gg \sqrt{\mathcal{E}^2 + \frac{\mathcal{B}_{\bot}^2}{D_{gs}}}
\end{equation}

 the eigenstates in this case are strong-axial field states $\ket{0}, \ket{S_z=+1}, \ket{S_z=-1}$ with corresponding eigenenergies $E_0, E_{-1} = D_{gs} - g\mu_BB_z, E_{+1} =  D_{gs} + g\mu_BB_z$. The fluctuations $\vec{\delta B_{dec}}$ and $\vec{\delta \Pi_{dec}}$ induce the fluctuations in the eigenergies (till the second order):

\begin{equation}
\label{eq:fluctuations_strong_field}
\begin{split}
& \delta E_{+1} = g\mu_B\delta B_{dec,z} + \frac{d_{\bot}^2(\delta\Pi_{dec,x'}^2 + \delta\Pi_{dec,y'}^2)}{2g\mu_BB_z} + \\
&+ \frac{1}{\sqrt{2}}\frac{(g\mu_B\delta B_{dec,x'})^2}{D_{gs} + g\mu_BB_z}+\frac{1}{\sqrt{2}}\frac{(g\mu_B\delta B_{dec,y'})^2}{D_{gs} - g\mu_BB_z}\\
& \delta E_{-1} = -g\mu_B\delta B_{dec,z} + \frac{d_{\bot}^2(\delta\Pi_{dec,x'}^2 + \delta\Pi_{dec,y'}^2)}{2g\mu_BB_z} + \\
&+ \frac{1}{\sqrt{2}}\frac{(g\mu_B\delta B_{dec,x'})^2}{D_{gs} + g\mu_BB_z}+\frac{1}{\sqrt{2}}\frac{(g\mu_B\delta B_{dec,y'})^2}{D_{gs} - g\mu_BB_z}\\
& \delta E_{0} = -\frac{(g\mu_B\delta B_{dec,x'})^2}{D_{gs} + g\mu_BB_z}-\frac{(g\mu_B\delta B_{dec,y'})^2}{D_{gs} - g\mu_BB_z}
\end{split}
\end{equation}

Considering the $\ket{0}\rightarrow\ket{-1}$ ODMR transition, the fluctuations of the resonance frequency are: 

\begin{equation}
\label{eq:fl_bis}
 \delta\nu_- = \delta E_{-1}  - \delta E_{0}  = g\mu_B\delta B_{dec,z} + \frac{d_{\bot}^2(\delta\Pi_{dec,x'}^2 + \delta\Pi_{dec,y'}^2)}{2g\mu_BB_z} 
\end{equation}

where we neglegted the term proportional to $\delta B_{dec,x'}, \delta B_{dec,y'} $ because $D_{gs}\gg g\mu_B\delta B_{dec,x'}, g\mu_B\delta B_{dec,y'}$.
In this scenario, the effect of fluctuations of the axial magnetic field $\delta B_z$ is at first order, instead the effects of the fluctuating total electric fields $\delta\Pi_x, \delta\Pi_y$ are at the second-order. This is due to the fact that, in this scenario $\langle S_z \rangle=\pm1$ for $\ket{S_z=\pm1}$. 

Following the same line of thought of the previous section, it can be shown that the FID decay is characterized by a coherence time

\begin{equation} 
\label{eq:coherence_time_strong_field}
T_{2,strong field}^* = \frac{1}{\sqrt{2}\pi}\frac{1}{\sqrt{(g\mu_B)^2\sigma_{B_{dec,z}}^2+\frac{\sigma_{\Pi_{dec,x}}^4+\sigma_{\Pi_{dec,y}}^4}{(2g\mu_BB_z)^2}}},
\end{equation}

In this scenario, the more effective source of decoherence is related to $B_{dec,z}$, i.e., the coupling with the different spin baths and the temporal fluctuations of the external magnetic field. 
 
\subsection{Limit case: Intermediate axial fields}

If we  consider the condition 

\begin{equation}
\mathcal{B}_{\parallel} \sim \sqrt{\mathcal{E}^2 + \frac{\mathcal{B}_{\bot}^2}{D_{gs}}}
\end{equation}

the eigenstates are partially dressed states $\ket{0}, \ket{-}_{\theta, \mathcal{B}_z} , \ket{+}_{\theta, \mathcal{B}_{\parallel}} $, see (Equation \ref{eq:eigenvectors_comb_B_z}). The fluctuations $\delta B_{dec_i}$ and $\delta \Pi_{dec,i}$ induce the fluctuations in the eigenergies (till the first order):

\begin{equation}
\label{eq:fluctuations_p_dressed}
\begin{split}
& \delta E_{+,\theta, \mathcal{B}_{\parallel}}= g\mu_B\delta B_{dec,z}\cos(\gamma) + d_{\bot}\delta\Pi_{dec,x'}\sin(\gamma) \\
& \delta E_{-,\theta, \mathcal{B}_{\parallel}}= -g\mu_B\delta B_{dec,z}\cos(\gamma) - d_{\bot}\delta\Pi_{dec,x'}\sin(\gamma) \\
& \delta E_{0} =-\frac{(g\mu_B\delta B_{dec,x'})^2}{D_{gs} + g\mu_BB_z}-\frac{(g\mu_B\delta B_{dec,y'})^2}{D_{gs} - g\mu_BB_z}
\end{split}
\end{equation}

where $\tan(\gamma)=\frac{E_{gap}/2}{g\mu_BB_{\parallel}}$ see (Eq.s (\ref{eq:eigenvalues_comb_B_z_bis}))

We underline that in Eq. \ref{eq:fluctuations_p_dressed}, the relative weight of $\delta B_{dec,z}$ and $\delta\Pi_{dec,y'}$ is set by the expectation value of $\langle S_z \rangle$ and 
$\langle S_{x}^2-S_{y}^2\rangle, \,\, \langle S_xS_y + S_yS_x \rangle$ on the eigenstates, e.g. $\langle +| S_z | +   \rangle_{B_z}=\cos(\gamma) $.

Considering the $\ket{0} \rightarrow \ket{-}_{\theta, B_{\parallel}}$ ODMR transition, the fluctuation of the resonance frequency is: 

\begin{widetext}
\begin{equation}
\label{eq:fl_ter}
 \delta\nu_- = \delta E_{-,\theta, \mathcal{B}_{\parallel}}  - \delta E_{0}  = -g\mu_B\delta B_{dec,z}\cos(\gamma) - d_{\bot}\delta\Pi_{dec,y'}\sin(\gamma) 
\end{equation}
\end{widetext}

Following the same line of thought of the previous subsections, it can be shown that the FID decay is characterized by a coherence time

\begin{widetext}
\begin{equation} 
\label{eq:coherence_time_p_dressed}
T_{2,part dressed}^* = \frac{1}{\sqrt{2}\pi}\frac{1}{\sqrt{(g\mu_B)^2\sigma_{B_{dec,z}}^2\cos(\gamma)+d_{\bot}^2\sigma_{\Pi_{dec,y'}}^2\sin(\gamma)}}.
\end{equation}
\end{widetext}

So, in this scenario, decoherence is due both to $B_{dec,z'}$ and to $\Pi_{dec,y'}$. The relative weights of the two different contributions depend on $\gamma$.

Summarizing for all three scenarios,:
\begin{itemize}
	\item The FID decay is a stretched exponential with p=2
	\item The FID decay is characterized by a coherence time $T_2^*$
	\item The value of the static fields defines the effectiveness of the different sources of decoherence
\end{itemize}

\section{Decoherence for an ensemble of NV centers}
\label{Calculation dec ensemble}
Let's consider now an ensemble of NV's. In general, we have to start from the expression of the FID decay for a single NV center in Eq. (\ref{eq:PLbis}) and average over the NV's ensemble. The FID decay depends on variance of the phase fluctuation, $\langle \delta \phi ^2\rangle$, and on the detuning $\Delta$ between the microwave excitation frequency $\nu_{MW}$ and the ODMR transition,  $\Delta= \nu_{MW}-\nu_{-,\theta, \mathcal{B}_{\parallel}}$. In principle, both $\langle \delta \phi ^2\rangle$ and $\nu_{-,\theta, \mathcal{B}_{\parallel}}$, and consequentely $\Delta$ take different values for the different spins of the ensemble following distributions $P(\langle \delta \phi ^2\rangle)$ and $f(\Delta)$. Let's considering two cases:

\begin{itemize}
	\item $\Delta$ does not vary over the ensemble. In this case we neglect gradients in the static fields $\vec{B}$ and $\vec{\Pi}$
	\item $\Delta$ varies over the ensemble. In this case we consider gradients in the static fields $\vec{B}$ and $\vec{\Pi}$
\end{itemize}

\subsection{$\Delta$ does not vary over the ensemble}

In this case we have $f(\Delta)=\delta(\Delta)$. The problem is to develop a suitable expression for $P(\langle \delta \phi ^2\rangle)$.  If we now consider the strong-axial-field scenario, and we neglect the effect due to fourth order fluctuations, $\langle \delta \phi ^2\rangle$ depends only on the variance of the fluctuating field $\sigma_{B-dec,z}$. If $\sigma_{B-dec,z}$ is mainly related to dipolar coupling, it can be shown that it is distributed as\cite{bauch2020decoherence}:

\begin{equation} 
\label{eq:B_dec_distribution}
P(\sigma_{B_{dec,z}}) =\frac{\sigma_{B_{dec,z-ens}}}{\sigma_{B_{dec,z}}^2}\sqrt{\frac{2}{\pi}}e^{-\sigma_{B_{dec,z ens}}^2/2\sigma_{B_{dec,z}}^2} .
\end{equation}

where $\sigma_{B_{dec,z-ens}}$ is related to the average coupling strength of the NV to the spin bath within the NV ensemble. For a detailed discussion of $P(\sigma_{B-dec,z})$ see Ref. \cite{dobrovitski2008decoherence}. This distribution has a maximum at $\sigma_{B_{dec,z}}=\frac{\sigma_{B_{dec,z-ens}}}{\sqrt{2}}$ and very heavy tails for higher values of $\sigma_{B_{dec,z}}$. 

Integrating $p_{FID}(\tau)$ over the distribution  $P(\sigma_{B_{dec,z}-ens})$, the ensemble averaged decay results:

\begin{equation} 
\label{eq:phi_ens_sf}
p_{FID}^{ens-strong-field}(\tau) = \frac{\left[1-e^{-\frac{\tau}{T_{2,ens, strong-field}^*}}\cos 2\pi \Delta \tau\right] }{2} .
\end{equation}

From the previous equation, it is clear that the ensemble-averaged decay is a simple exponential, $p=1$ with 

\begin{equation}
T_{2,ens, strong-field}^*=\frac{1}{g\mu_B\sigma_{B_{dec,z-ens}}}
\end{equation}
If we now consider the dressed states scenario, and we neglect the effect due to fourth-order fluctuations, $\langle \delta \phi ^2\rangle$ depends only on  $\sigma_{\Pi_{dec,x'}}$. 

If we consider $\sigma_{\Pi_{dec,x'}}$ as mainly related to temporal fluctuation, we can consider that it  does not vary over the NV's ensemble, formally  $P(\sigma_{\Pi_{dec,x'}})=\delta(\sigma_{\Pi_{dec,x'}} - \sigma_{\Pi_{dec,x'-ens}})$, involving a Dirac $\delta$ function.   

 Integrating $p_{FID}(\tau)$ over the distribution  $P(\sigma_{\Pi_{dec,y'}})$, the ensemble averaged decay results: \\
\begin{widetext}
\begin{equation} 
\label{eq:phi_ens_dressed}
\begin{split}
 p_{FID}^{ens-dressed}(\tau) &= \frac{\left[1-\int e^{-d_{\bot}^2\sigma_{\Pi_{dec,x'}}^2\tau^2}\cos 2\pi \Delta \tau \delta(\sigma_{\Pi_{dec,x'}} - \sigma_{\Pi_{dec,x'-ens}}) d\sigma_{\Pi_{dec,x'}} \right]}{2}=\\
 &= \frac{\left[1-e^{-\left(\frac{\tau}{T_{2,ens, dressed}^*}\right)^2}\cos 2\pi \Delta \tau \right]}{2}\\
\end{split}
\end{equation}
\end{widetext}
The ensemble-averaged decay is a stretched exponential (p=2) with 
\begin{equation}
T_{2,ens, dressed}^*=\frac{1}{d_{\bot}\sigma_{\Pi_{dec,x'-ens}}}.
\end{equation}
If we now consider the partially dressed states scenario, $\langle \delta \phi ^2\rangle$ depends only both on  $\sigma_{\Pi-dec,y'}$ and $\sigma_{B_{dec,z}}$. If we use $P(\sigma_{B_{dec,z}})$ and $P(\sigma_{\Pi_{dec,x'}})$ defined before, the ensemble averaged decay takes the form
\begin{widetext}    
\begin{equation} 
\label{eq:phi_ens_p_dressed}
\begin{split}
p_{FID}^{ens-part-dressed}(\tau) &= \frac{\left[1-\int e^{(-\cos(\gamma)g^2\mu_B^2\sigma_{B_{dec,z}}-\sin(\gamma)d_{\bot}^2\sigma_{\Pi_{dec,y}}^2)\tau^2} \cos (2\pi \Delta \tau) P(\sigma_{B_{dec,z}})P(\sigma_{\Pi_{dec,y'}}) d\sigma_{\Pi_{dec,y'}} d \sigma_{B_{dec,z}}\right]}{2}=\\
& = \frac{\left[1-e^{-\cos(\gamma)\frac{\tau}{T_{2,ens, strong-field}^*}}e^{-\sin(\gamma)\left(\frac{\tau}{T_{2,ens, dressed}^*}\right)^2}\cos 2\pi \Delta \tau \right]}{2}\\
\end{split}
\end{equation}
\end{widetext}
The ensemble-averaged decay is a product of a simple exponential with a stretched exponential. The relative weight of the two exponentials is set by $\gamma$ with $\tan(\gamma)=\frac{E_{gap}/2}{g\mu_BB_z}$.\\
Summarizing, in the absence of field gradients, at $B_z=0$, the decay will be close to a stretched exponential with p=2; for high axial field  $B_z$, the decay will be close to a simple exponential; for intermediate values of $B_z$ the decay is a product of a simple exponential and a stretched exponential.

\subsection{$\Delta$ varies over the ensemble}

Let's start neglecting the decoherence for the single NV. In this case, $\langle \delta \phi ^2\rangle=0$ for every NV center in the ensemble, i.e. $P(\langle \delta \phi ^2\rangle)=\delta(\langle \delta \phi ^2\rangle)$. Instead, we consider that the static field $\vec{B}$ and $\vec{\Pi_y'}$ varies over the ensemble, inducing a distribution $P(\Delta)$ in the detuning $\Delta$.  The ensemble-averaged decay takes the form:
\begin{equation} 
\label{eq:phi_gradient}
\begin{split}
& p_{FID}^{ens-gradient-sf}(\tau) =\\
& =\frac{\left[1-1/2\left(\int e^{2\pi i \Delta \tau }P(\Delta) d \Delta+\int e^{-2\pi i \Delta \tau }P(\Delta) d \Delta\right) \right]} {2}\\
& =\frac{\left[1-1/2(\tilde{P}(\tau)+\tilde{P}(\tau)^*)\right]} {2}\\
\end{split}
\end{equation}

The ensemble-averaged decay involves the Fourier transform  $\tilde{P}(\tau)$ of the distribution of detuning $P(\Delta)$. 

If we now consider the strong-axial-field scenario, from Eq. (\ref{eq:phi_ens_sf}), we have: 
\begin{equation} 
\label{eq:phi_gradient_sf}
\begin{split}
& p_{FID}^{ens-gradient-dressed}(\tau) =\\
& =\frac{\left[1-\int e^{-\frac{\tau}{T_{2,ens, strong-field}^*}} \cos 2\pi \Delta \tau P(\Delta) d \Delta\right]} {2}\\
& =\frac{\left[1- e^{-\frac{\tau}{T_{2,ens, strong-field}^*}}(\tilde{P}(\tau)+\tilde{P}(\tau)^*)\right]} {2}
\end{split}
\end{equation}
where we use the fact that $T_{2,ens, strong-field}^* \neq T_{2,ens, strong-field}^*(\Delta)$. The ensemble-averaged decay involves the product of the pure exponential with p=1 and the Fourier transform  $\tilde{P}(\tau)$ of the distribution of detuning $P(\Delta)$.\\
If we now consider the dressed state scenario, from Eq. (\ref{eq:phi_ens_dressed}), we have: 
\begin{equation} 
\label{eq:phi_gradient_dressed}
\begin{split}
& p_{FID}^{ens-gradient-dressed}(\tau) =\\
& =\frac{\left[1-\int e^{-\left(\frac{\tau}{T_{2,ens, dressed}^*}\right)^2} \cos 2\pi \Delta \tau P(\Delta) d \Delta\right]} {2}\\
& =\frac{\left[1- e^{-\left(\frac{\tau}{T_{2,ens, dressed}^*}\right)^2}(\tilde{P}(\tau)+\tilde{P}(\tau)^*)\right]} {2}\\
\end{split}
\end{equation}

where we use the fact that $T_{2,ens, dressed}^* \neq T_{2,ens, dressed}^*(\Delta)$. The ensemble-averaged decay involves the product of the stretched exponential with p=2 and the Fourier transform  $\tilde{P}(\tau)$ of the distribution of detuning $P(\Delta)$.

If we now consider the partially-dressed states scenario, from Eq. (\ref{eq:phi_ens_p_dressed}), we have:
\begin{widetext}
\begin{equation} 
\label{eq:phi_gradient_p_dressed}
\begin{split}
& p_{FID}^{ens-gradient-p-dressed}(\tau) =\\
& =\frac{\left[1-\int e^{-\cos(\gamma)\frac{\tau}{T_{2,ens, strong-field}^*}}e^{-\sin(\gamma)\left(\frac{\tau}{T_{2,ens, dressed}^*}\right)^2} \cos 2\pi \Delta \tau P(\Delta) d \Delta\right]} {2}\\
& =\frac{\left[1- e^{-\cos(\gamma)\frac{\tau}{T_{2,ens, strong-field}^*}}e^{-\sin(\gamma)\left(\frac{\tau}{T_{2,ens, dressed}^*}\right)^2}(\tilde{P}(\tau)+\tilde{P}(\tau)^*)\right]} {2}
\end{split}
\end{equation}
\end{widetext}
The ensemble-averaged decay involves the product of the simple exponential, the stretched exponential with p=2, and the Fourier transform  $\widehat{P}(\tau)$ of the distribution of detuning $\Delta$.\\ In summary, gradients in the static field change the time dependence of the decay. In particular, for a strong-axial-field scenario, it is no longer a simple exponential, and for a dressed-state scenario, it is no longer a stretched exponential with p=2.

\section{Free Induction Decay measurements for strong axial field states}
\label{Calculations High Field}

In Figure \ref{fig:FID_High_Field}, FID measurements recorded for the $NV_{2}$ family are reported, see Fig. \ref{fig:CW_ODMR_spectrum}. The $B_{\parallel}$ component can be calculated from the full range CW-ODMR spectrum in Fig. \ref{fig:CW_ODMR_spectrum} yielding $B_{\parallel} \approx 3  mT$. 
\begin{figure}[h!]
	\includegraphics[width=\columnwidth]{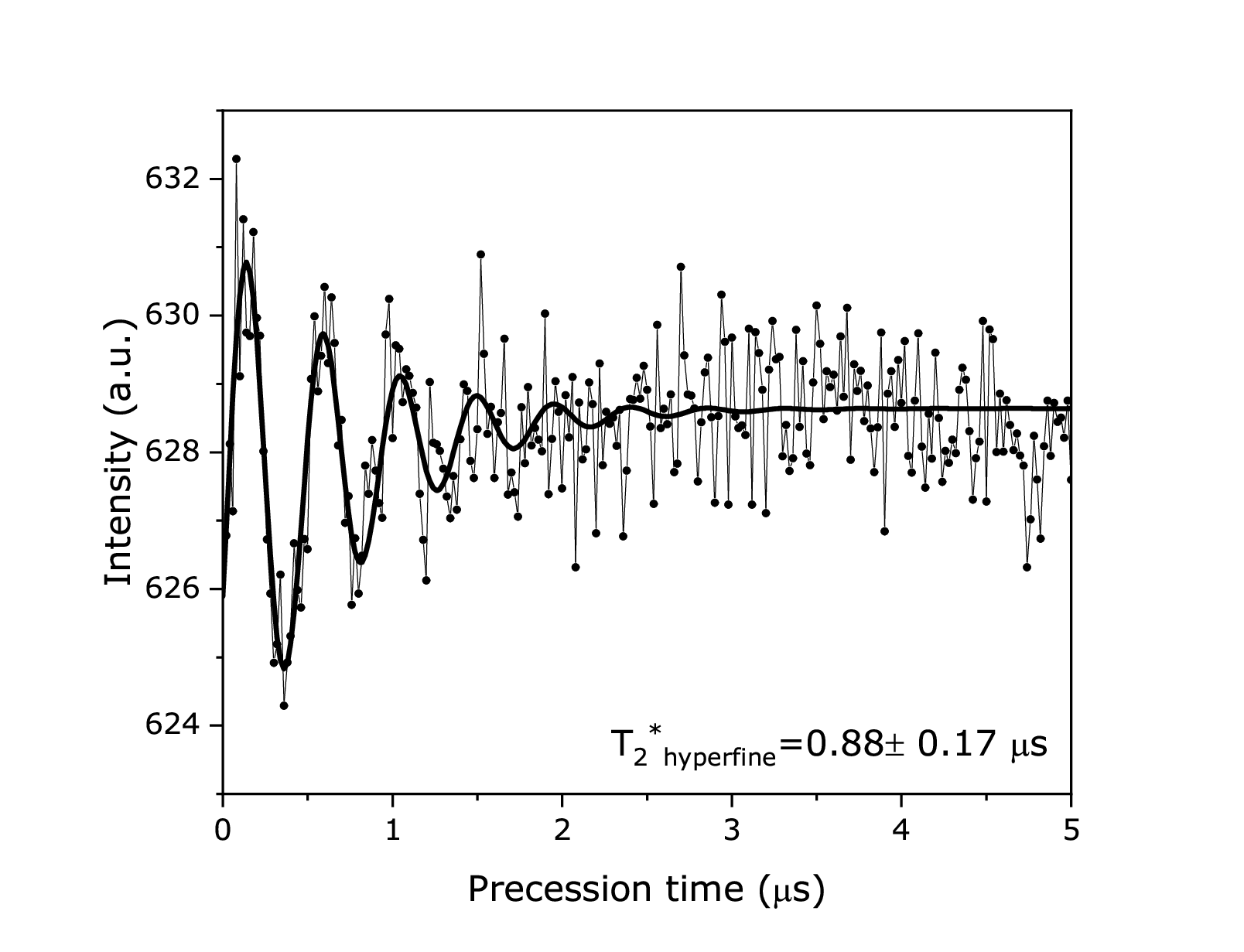}
	\caption{FID relaxation for a microwave on resonance with the central peak of a hyperfine family at high fields}
	\label{fig:FID_High_Field} 
\end{figure}

The shorter $T^{*}_{2}$ indicates the high-field character of the state (the complete list of $T^{*}_{2}$ values can be found in Table 1 in the main text). The FID data was recorded by tuning the MW frequency in resonance with the central hyperfine peak, therefore only a single detuning ($\nu=2.16  \,  MHz$) is visible in the graph. As described in the previous section, strong-axial field states couple with the spin bath and fluctuations of external magnetic fields, leading to shorter coherence times with respect to dressed states, where the largest sources of decoherence are electric fields and strain fluctuations.

\begin{figure}[h!]
	\includegraphics[width=\columnwidth]{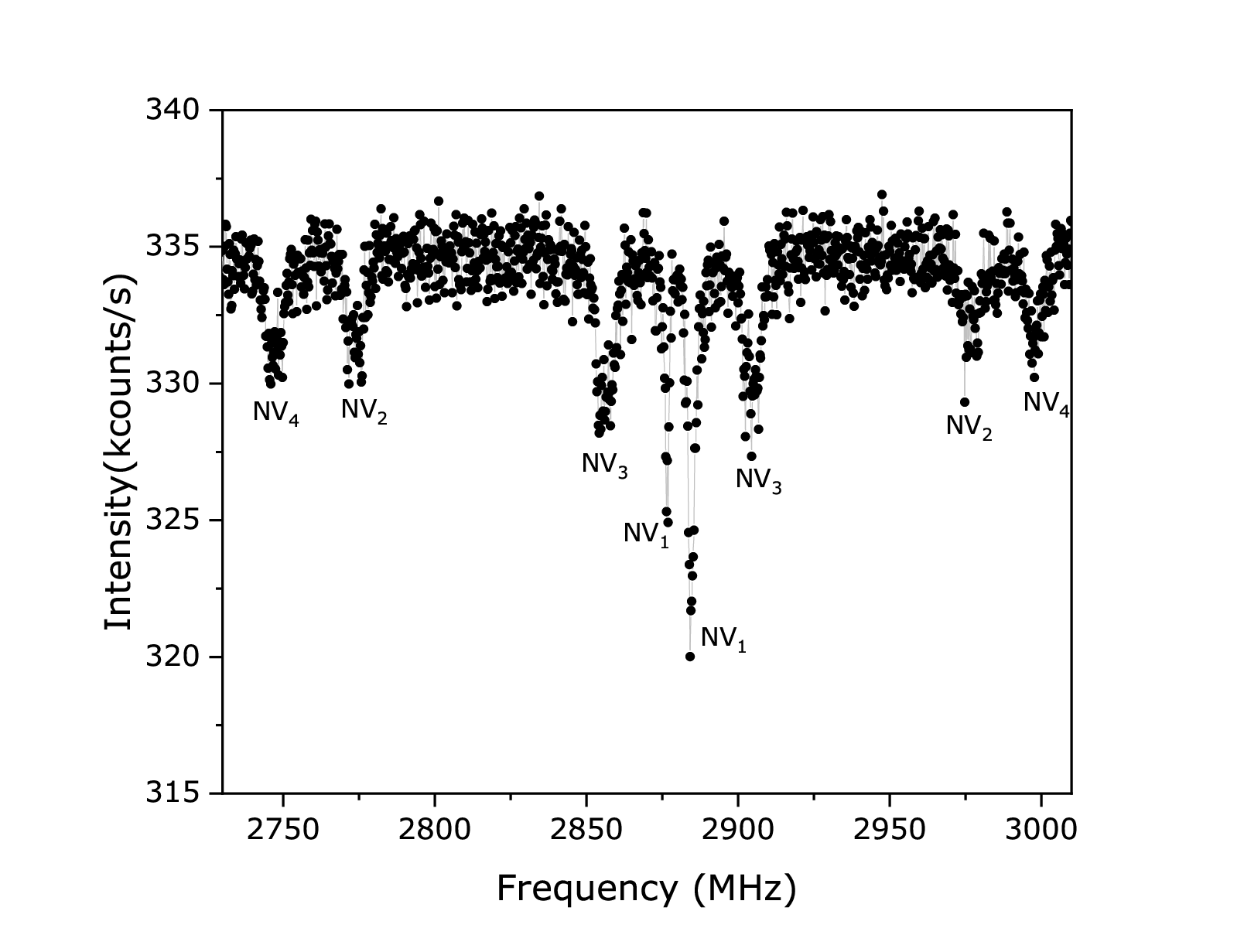}
	\caption{{CW-ODMR spectrum with identification of the crystallographic orientation}}
	\label{fig:CW_ODMR_spectrum} 
\end{figure}

\clearpage

\end{document}


\title{Interplay between dressed and  strong-axial field states in solid state spin ensembles-Supplementary Materials}

\author{G. Zanelli$^{1,2}$}
\author{E. Moreva$^1$}
\author{E. Bernardi$^1$}
\email{e.bernardi@inrim.it}
\author{E. Losero$^1$}
\author{S. Ditalia Tchernij$^{2,3}$}
\author{J. Forneris$^{2,3}$}
\author{\v{Z}. Pastuovi\'{c}$^4$}
\author{P. Traina$^1$}
\author{I. P. Degiovanni$^1$}
\author{M. Genovese$^{1,3}$}

\affiliation{$^1$Istituto Nazionale di Ricerca Metrologica, Strada delle cacce 91, Turin, Italy}\email{e.bernardi@inrim.it}
\affiliation{$^2$Physics Department and NIS Centre of Excellence - University of Torino, Torino, Italy}
\affiliation{$^3$Istituto Nazionale di Fisica Nucleare (INFN) Sez. Torino, Torino, Italy}
\affiliation{$^4$Centre for Accelerator Science, Australian Nuclear Science and Technology Organisation, New Illawarra rd., Lucas Heights, NSW 2234, Australia}

\widetext

\maketitle

\newcommand{\ket}[1]{\mbox{\ensuremath{|#1\rangle}}}
\newcommand{\bra}[1]{\mbox{\ensuremath{\langle#1|}}}
\renewcommand{\theequation}{S\arabic{equation}}
\renewcommand{\thefigure}{S\arabic{figure}}
\renewcommand{\thetable}{S\arabic{table}}


\section{Experimental Set-up}

\subsection{Experimental Setup}
The experimental setup is a single-photon sensitive confocal microscope in an inverted configuration. The excitation source is a 532 nm CW-laser (Prometheus 100NE), which can be pulsed using an acousto-optic modulator (Gooche \& Housego 3080 – 125). The excitation is focused down to $\sim 2 \mu m $ spot diameter onto the sample using an air objective (Olympus UPLANFL, 60x NA=0.67), which also collects the NV photoluminescence (PL) signal.  The emitted PL is filtered with a 650 nm long-pass filter, and 10\% of the emitted intensity is collected by a multi-mode optical fiber (core diameter = 50 $\mu$m) and recorded by a single photon avalanche photodiode (SPAD). The signal from the detector is recorded by a digital counter (Swabian Instruments Time Tagger 20) and sent to the software used for instrument control \cite{binder2017qudi}. The MW signal required for ODMR is generated by an analog microwave generator (Keysight N5183B), then supplied to the power amplifier (SHL-16W-43-S+) using a switch (ZASW-2-50DRA+), which can be employed to pulse the signal using a pulse generator (Swabian Instrument Pulse Streamer 8/2). Finally, the MW arrives at the sample exploiting a planar antenna optimized to work efficiently around 2.87 GHz \cite{sasaki2016broadband}. The bias magnetic field is supplied by a small magnet placed using a micro-manipulator and adjusted using a set of 3 Helmholtz coils. The sample is a $3 \times 3 \times 0.5$ mm CVD type IIa single-crystal diamond ($\text{N} < 1 \; \text{ppm}, \text{B} < 0.05 \; \text{ppm}$) from Element Six. NV centers have been created by 10 keV ${}^{14}\text{N}^{+}$ ion implantation with F=$1\cdot10^{14}$ cm$^{-2}$ followed by a 2h thermal annealing at 950°C. This process yields a final concentration of NVs = $3\cdot 10^{18} cm^{-3}$ (17 ppm) at a 10 nm depth from the surface while the concentration of residual substitutional nitrogen in the implanted region is $5.8\cdot 10^{19}$ cm$^{-3}$ (330 ppm).

\begin{figure*}[h!]
	\centering
	\includegraphics[width=0.75\textwidth]{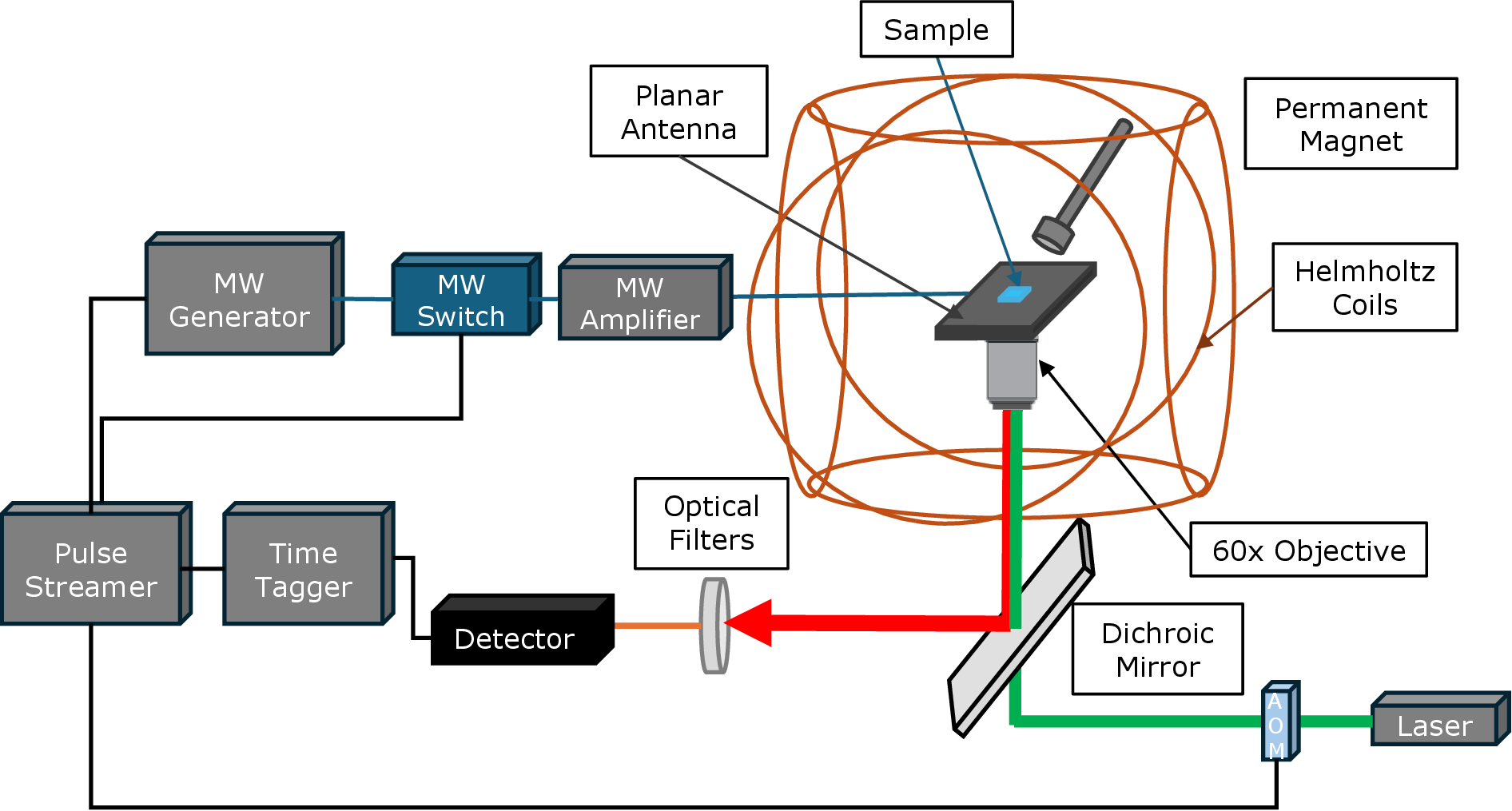}
	\captionof{figure}{Experimental Setup}\label{setup}
\end{figure*}

\subsection{CW-ODMR and FID measurements parameters}
\begin{figure}[h!]
	\centering
	\begin{subfigure}[b]{0.4\textwidth}
		\centering
		\includegraphics[height=3.5 cm]{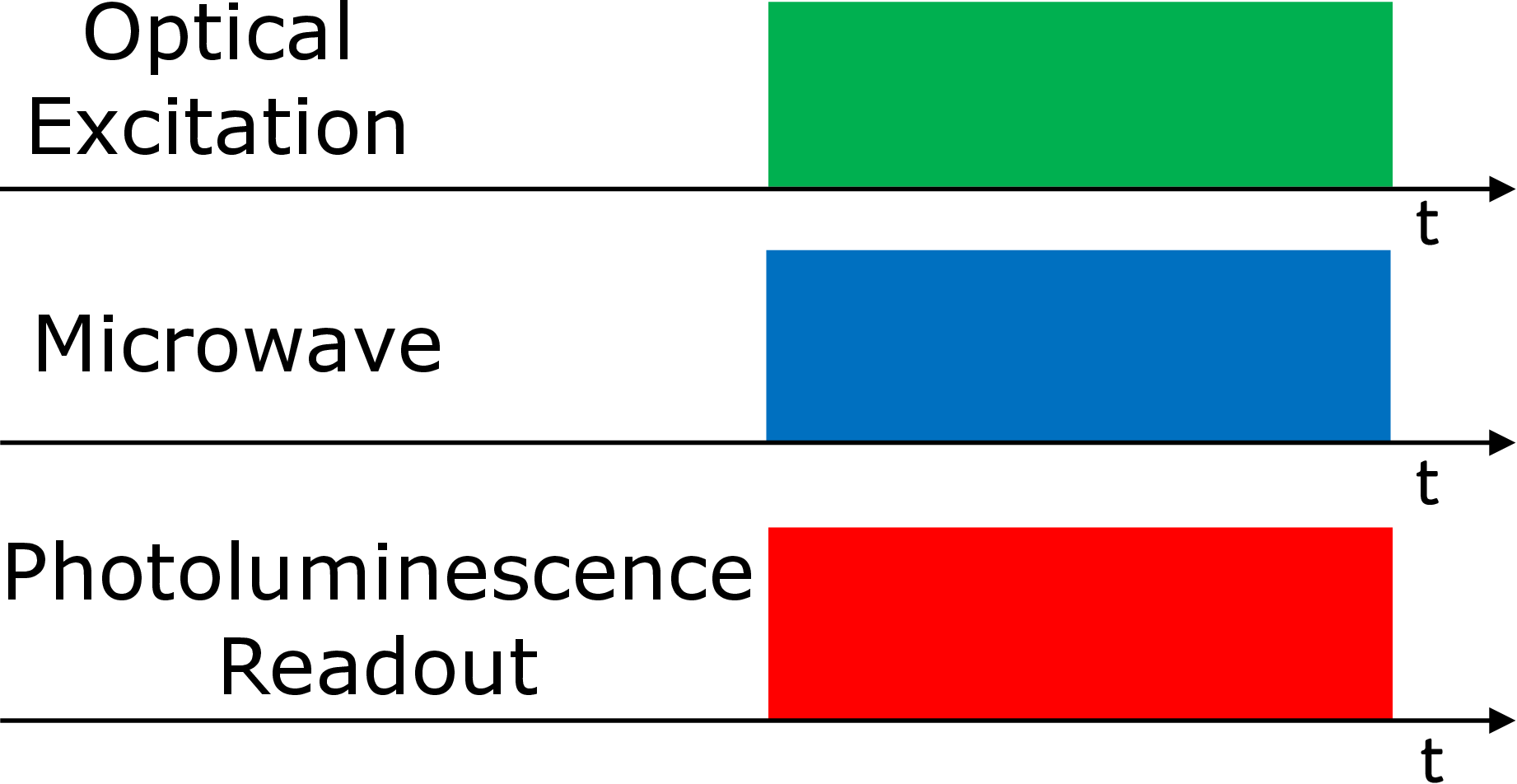}
		\caption{CW-ODMR scheme}
		\label{fig:CW-ODMR sequence}
	\end{subfigure}
	\hfill
	\begin{subfigure}[b]{0.55\textwidth}
		\centering
		\includegraphics[height= 3.5 cm]{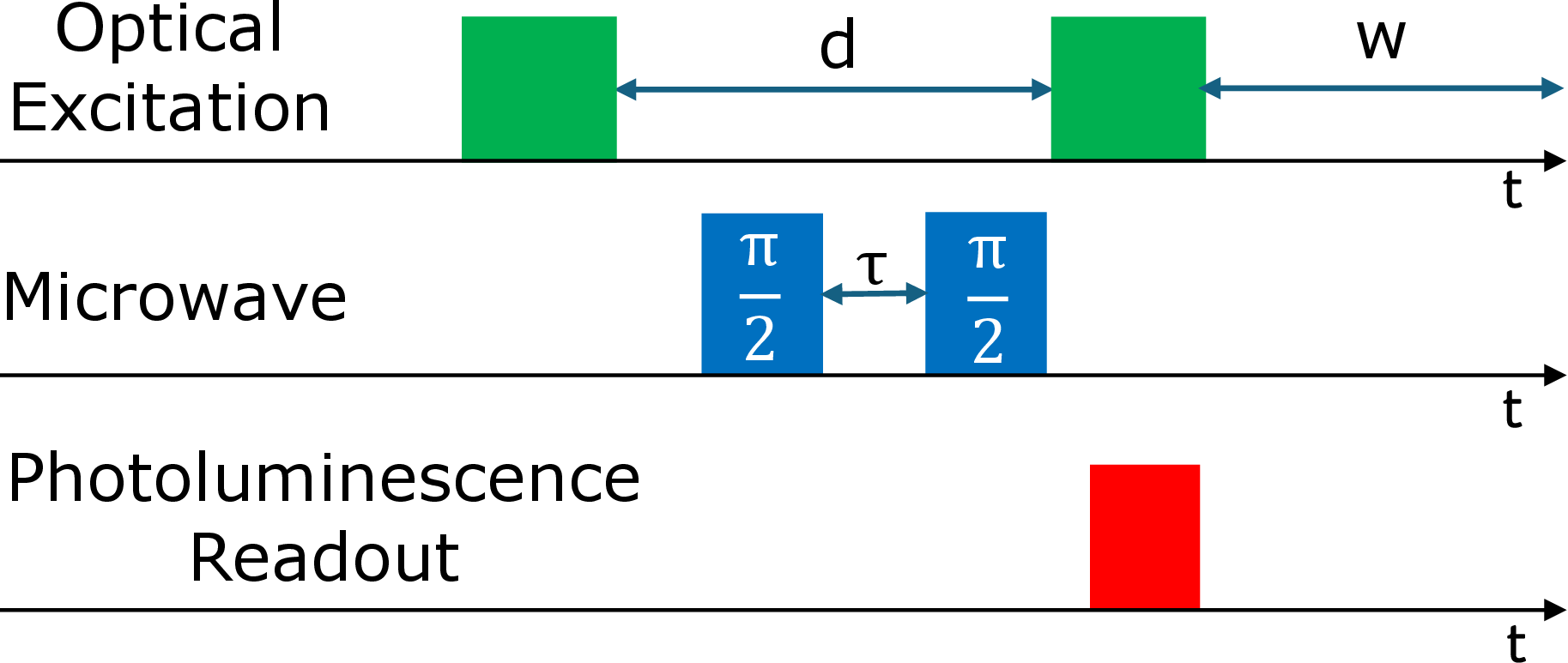}
		\caption{FID sequence}
		\label{fig:Ramsey_scheme}
	\end{subfigure}
	\caption{(a) CW-Optically Detected Magnetic Resonance(ODMR) sequence (b) Free Induction Decay (FID) Sequence}
	\label{fig:CW_vs_FID}
\end{figure}

Sample characterization has been carried out by using two main techniques, namely, Continuous-Wave Optically Detected Magnetic Resonance (CW-ODMR) and Free Induction Decay (FID). The two techniques are based on the optical spin polarization induced on the NV center by laser irradiation (532 nm) and the application of a MW field to induce transitions between different spin sub-levels in the ground state. CW-ODMR is characterized by a continuous application of initialization, driving and readout in the form of optical and microwave fields and presents an overall easier experimental implementation (Fig.\ref{fig:CW-ODMR sequence}). FID technique presents a sequence of pulses for initialization, driving and readout, resulting in a  more challenging experimental implementation. The CW-ODMR measurements are performed by applying  5 mW of CW laser light, and 30 mW of CW microwave power for 10 minutes of measurement time. FID measurements required 120 minutes to acquire sufficient photon statistics. FID measurements parameters are reported in Table S.1, while the optical power is 20 mW and the microwave power is 10 W.
The selection of initialization and readout pulse duration is strictly related to the optical power, as larger optical powers induce faster dynamics in the NV center \cite{giri2018relaxometry}. Additionally, the time delay d between initialization and readout laser pulses is kept constant to avoid drifts in the PL signal as the time $\tau$ between the MW pulses increases.

\begin{table}[h!]
	\centering
	\begin{tabular}{@{} l S[table-format=3.0] S[table-format=5.0] @{}}
		\toprule
		{Stage Duration ($\mu s$) } & {CW-ODMR} & {FID} \\
		\midrule
		Optical Initialization & $\infty$ & {2.0} \\
		Photoluminesce Readout        & $\infty$ & {1.5} \\
		Waiting time ($w$)  & {--\,--} &  {3}\\
		Delay time  ($d$)    & {--\,--} & {7.5}\\
		\bottomrule
	\end{tabular}
	\caption{Sequence sections durations}
\end{table}

\section{Calculations of total electric field inside bulk diamond}
The simulation of the total electric field is conducted to obtain information about  its orientation  with respect to the orthogonal field. This goal is achieved into two main steps: first, the calculation of the orientation of magnetic field performed using 3 orientations of NVs in high-field configuration, then, the orientation of the total electric field with respect to the orthogonal field is computed by monitoring the effect on the NV orientation in orthogonal-field configuration. 

\subsection{Orientation of orthogonal field}
The orientation and intensity of a 3D magnetic field can be obtained via vector magnetometry using an ensemble of NV centers. To perform this process, it is necessary to know the four orientations in the space of the different NV families, knowing that the [100] sample orientation corresponds with the laboratory z-axis. The labels assigned to each family are in Table \ref{tab:families}. \\
The labels are then assigned to each CW-ODMR peak by recording subsequent spectra after adding a small axial field component ($B_{\parallel}\approx 0.5 mT$) in each of the four different orientations respectively. The behavior of the spectroscopic lines can be easily computed via the Electron Zeeman term of the complete Hamiltonian:

\begin{gather*}
	H_{Zeeman}=-g_{e}\mu_{B}S\cdot B 
\end{gather*}
\text{where:} \\
\begin{align*}
g_{e}: \text{electron gyromagnetic ratio} \quad
\mu_{B}: \text{Bohr magneton} \\
S: \text{total spin} \quad
B: \text{total magnetic field}
\end{align*}

By monitoring the two peaks that exhibit the largest shifts at each step, it is possible to assign a crystallographic orientation to each peak, see Fig.6 in the main text:

\begin{table}[h!]
	\centering
	\begin{tabular}{@{} l S[table-format=2.0] S[table-format=5.0] @{}}
		\toprule
		{NV Label } & {Crystallographic Orientation}  \\
		\midrule
		$NV_{1}$	&	{[\,\,1\,\,1\,\,1\,\,]}  \\
		$NV_{2}$	&	{[\,\,1-1-1\,\,]}  \\
		$NV_{3}$ 	&	{[-1\,\,1-1\,\,]}  \\
		$NV_{4}$	&	{[-1-1\,\,1\,\,]}  \\
		\bottomrule
	\end{tabular}
	\caption{Crystallographic orientation of the different NV families}
    \label{tab:families}
\end{table}

Once the peaks are labeled, it is possible to obtain the direction of the orthogonal magnetic field in the reference frame of the $NV_{1}$ family. This is acquired by simulating the effect of a bias magnetic field rotating on the [111] plane and then selecting the value where the spectroscopy lines of the $NV_{2-3-4}$ families are ordered as in Fig.6 of the main text. Finally, the magnitude of the field is adjusted to fit the computed spectroscopy lines with the measured ones, yielding:

\begin{gather*}
	B_{\perp} = [\;3.89\;3.27\;\;0]\, \text{mT} \\
	|B_{\perp}|=5.08\, \text{mT}, \ \phi_{B_{\perp}}=40^\circ
\end{gather*}

\begin{figure}[hb!]
	\includegraphics[width= 14 cm]{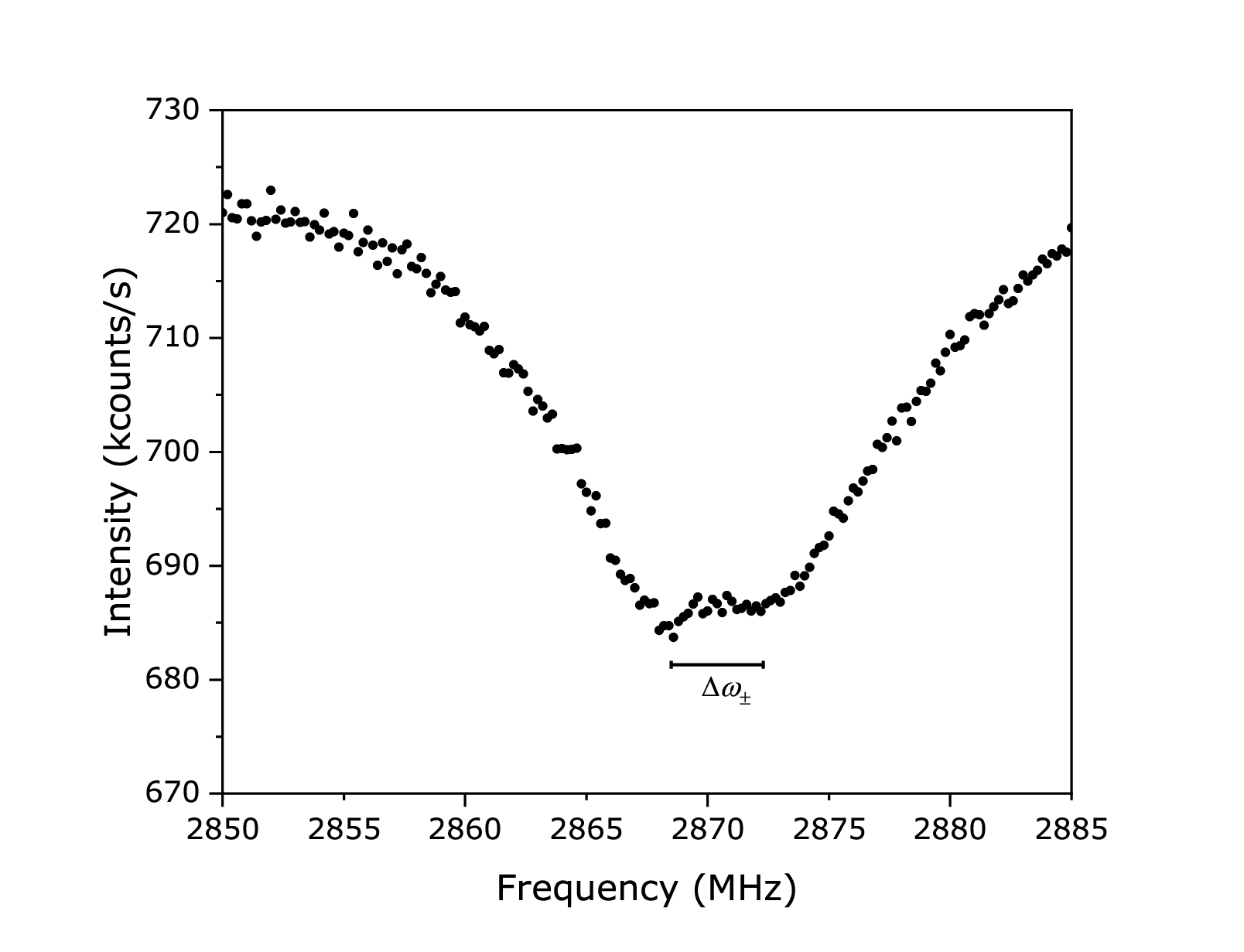}
	\captionof{figure}{Zero Field CW-ODMR}\label{ZFCW}
\end{figure}

\subsection{Computation of total electric field components}
The computation of the total electric field components in the NV Hamiltonian is not straightforward. However, if the orientation of the bias magnetic field is known, it is possible to reconstruct the total electric field using the lines from the $NV_{1}$ family (orthogonal field configuration), given the total electric field component Hamiltonian that is:

\begin{align*}
	&H_{\pi}=- d_{\bot}\left[\Pi_x(S_x^2- S_y^2)-\Pi_y(S_xS_y + S_yS_x)\right] +d_{\parallel}[S^{2}_{z}-\frac{1}{3}S(S+1)]
\end{align*}
where:\\

\[\frac{d_{\perp}}{\hbar}=17 \, \text{Hz} \, \text{V}^{-1} \, \text{cm},\ \frac{d_{\parallel}}{\hbar}=0.35\text{Hz} \, \text{V}^{-1} \, \text{cm}\]

Due to the larger coupling strength $(d_{\perp} \gg d_{\parallel})$, only the $\Pi_{x}$ and $\Pi_{y}$ components of the total electric field are considered. The frequency splitting from the zero field CW-ODMR gives information on the magnitude of $\Pi_{x}$ and $\Pi_{y}$ see Ref. \cite{jamonneau2016competition}, i.e.:
\begin{gather*}
\Delta \omega_{\pm}=2\sqrt{\Pi^{2}_{x}+\Pi^{2}_{y}}\approx 5\, \text{MHz}
\end{gather*}
The value obtained will serve as a constraint for evaluating the magnitude of the total electric field.

\begin{figure}[h!]
	\includegraphics[width=0.7\textwidth]{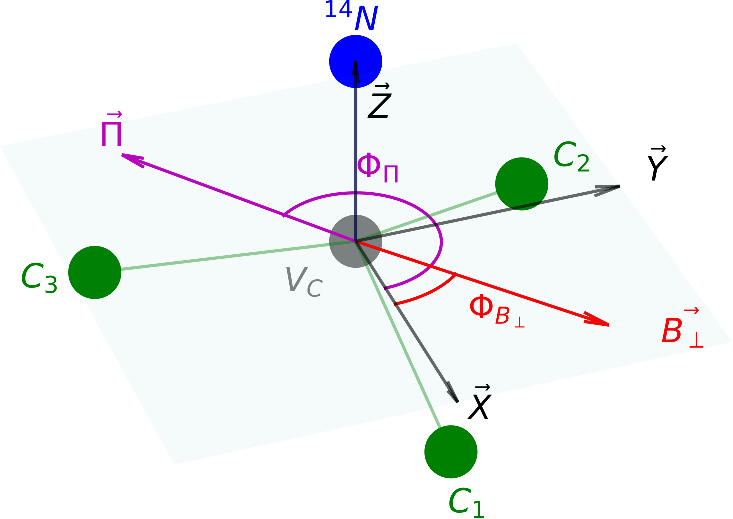}
	\captionof{figure}{Orthogonal field $\vec{B_{\perp}}$ and strain $\vec{\Pi}$ relative orientations}\label{B_orth}
\end{figure}

Starting from this, the total electric field values obtained are: $\Pi_{x} = -124000 \, \text{V}^{-1} \, \text{cm}^{-1}, \Pi_{y} = -94000 \, \text{V}^{-1} \, \text{cm}^{-1},$.
Finally, the relative angle between the total electric field and the magnetic orthogonal field is obtained from the complete numerical simulation as: $\phi_{\mathcal{E}}-\phi_{\mathcal{B}} = 177.2^{\circ} $, corresponding to $\phi_{\mathcal{E}}=217.2^{\circ}$. In Fig. \ref{B_orth}, the relative orientations of strain and orthogonal magnetic field are represented.
\clearpage